\documentclass[reprint,aps,prb,superscriptaddress,floatfix]{revtex4-1}
\usepackage[utf8]{inputenc}
\usepackage{amssymb,amsmath}
\usepackage{siunitx}
\usepackage{graphicx}
\usepackage{tabularx}
\usepackage{float}
\usepackage{dcolumn}
\usepackage{multirow}
\usepackage{color}
\usepackage[english]{babel}
\usepackage{hyperref}
\usepackage{cases}
\usepackage{bm}
\usepackage{enumitem}
\usepackage{comment}
\graphicspath{{Images/}{ImagesSupplemental/}}

\begin{document}
	\title{Lifting of Spin Blockade by Charged Impurities 
		in Si-MOS Double Quantum Dot Devices}
	\author{Cameron King}
	\email{cameron.eric.king@gmail.com}
	\affiliation{Department of Physics, University of Wisconsin-Madison, 
		Madison, Wisconsin 53706, USA}
	\author{Joshua S.\ Schoenfield}
	\affiliation{Department of Physics and Astronomy, University of California 
		Los Angeles, Los Angeles, California 90095, USA}
	\author{M.J.\ Calder\'{o}n}
	\affiliation{Instituto de Ciencia de Materiales de Madrid, ICMM-CSIC, 
		Cantoblanco, E-28049 Madrid (Spain)}
	\author{Belita Koiller}
	\affiliation{Instituto de F\'{i}sica, Universidade Federal do Rio de Janeiro,
		C.P.\ 68528, 21941-972 Rio de Janeiro, Brazil}
	\author{Andr\'{e} Saraiva}
	\affiliation{Instituto de F\'{i}sica, Universidade Federal do Rio de Janeiro,
		C.P.\ 68528, 21941-972 Rio de Janeiro, Brazil}
\affiliation{School of Electrical Engineering and Telecommunications, The University of New South Wales, Sydney, NSW 2052, Australia}
	\author{Xuedong Hu}
	\affiliation{Department of Physics, University at Buffalo, SUNY, 
		Buffalo, New York 14260-1500, USA}
	\author{HongWen Jiang}
	\affiliation{Department of Physics and Astronomy, University of California 
		Los Angeles, Los Angeles, California 90095, USA}
	\author{Mark Friesen}
	\email{friesen@physics.wisc.edu}
	\affiliation{Department of Physics, University of Wisconsin-Madison, 
		Madison, Wisconsin 53706, USA}
	\author{S.\ N.\ Coppersmith}
	\email{snc@physics.wisc.edu}
	\affiliation{Department of Physics, University of Wisconsin-Madison, 
		Madison, Wisconsin 53706, USA}
	\affiliation{School of Physics, The University of New South Wales, Sydney NSW 2052, Australia}
	
	\date{\today}
	
	\begin{abstract}
		One obstacle that has slowed the development of electrically gated 
		metal-oxide-semiconductor (MOS) singlet-triplet qubits is the frequent lack 
		of observed spin blockade, even in samples with large singlet-triplet 
		energy splittings. We present theoretical and experimental evidence that 
		this problem in MOS double quantum dots can be caused by stray positive 
		charges in the oxide inducing accidental localized levels near the device’s active 
		region that lead to the lifting of the spin blockade. 
		We also present evidence that 
		these effects can be mitigated by device design modifications, such as 
		overlapping gates.
	\end{abstract}
	
	\pacs{73.21La, 73.63.Kv, 72.25.-b, 85.35.Gv}
	
	\maketitle
	
\section{Introduction}
\label{introduction}
Silicon metal-oxide-semiconductor (Si-MOS) devices form the foundation of 
current electronics, and the manufacturability, reliability, and scalability 
of MOS technology are attractive reasons to develop MOS devices for quantum
information processing~\cite{Loss:1998ia,Zwanenburg:2013gla}. 
Spin coherence times in silicon can be quite long~\cite{Tyryshkin:2012fi}, 
enabling the demonstration of high-fidelity quantum dot spin qubits in MOS 
devices~\cite{Veldhorst:2014eq, Veldhorst:2015je, Jock:2018cu}.
Given these successes, MOS is a natural architecture for further development of 
electrically controlled spin qubits,
such as singlet-triplet qubits~\cite{Levy:2002ex, Petta:2005kn}, 
which have recently been demonstrated~\cite{Jock:2018cu}. 

\textcolor{black}{Pauli spin blockade arises as a consequence of spin  
conservation in electron tunneling~\cite{Ciorga:2000br, Ono:2002gz}: when the  
singlet-triplet splitting is large in a quantum dot, a $(1,1)$ spin triplet  
cannot transition to the $(2,0)$ configuration while a $(1,1)$ singlet 
can~\cite{Ono:2002gz, Hao:2014ea}
(here, $(m,n)$ denotes $m$ electrons in the left quantum dot and 
$n$ electrons in the right dot).
Because of spin blockade, the electron spin and charge configurations are 
correlated, which can be used to initialize and detect the state of a spin 
qubit, particularly the singlet-triplet qubit~\cite{Petta:2005kn, 
Maune:2012iu,Wu:2014fz,HarveyCollard:2017ic}.
As such, spin blockade is a crucial ingredient for spin-based quantum computing 
architectures.}

\textcolor{black}{Spin blockade requires a large singlet-triplet splitting in the 
detection dot and a non-magnetic environment so that singlet and triplet 
electron spin states have long lifetimes. 
A Si-MOS quantum dot typically has a large conduction band valley splitting 
and a large singlet-triplet splitting, and isotopic enrichment helps suppress 
magnetic noise from nuclear spins.
One would therefore expect that a Si-MOS double quantum dot should provide a 
favorable environment for spin blockade, and indeed spin blockade has been 
observed experimentally~\cite{Hao:2014ea, Veldhorst:2017ht,  
HarveyCollard:2017ic, Xiao:2010cx, Maurand:2016cj}.
However, despite massive efforts within the research community 
to remove blockade-lifting mechanisms such as low valley splitting and 
nuclear spins, many samples with large and positive singlet-triplet splittings
(hundreds of 
\si{\micro\electronvolt}, as measured using excited-state spectroscopy) 
fail to exhibit spin blockade. 
In the experiments reported here, nine out of ten samples, all with
large singlet-triplet splittings, failed to exhibit spin blockade.
}

Here we show that the absence of spin blockade in a MOS device can be explained 
via the presence of an unintentional level in the system containing an electron 
that is exchange coupled to the gate-defined dots. 
We present calculations demonstrating that these levels are likely induced by 
charges present within the oxide layer in typical samples. 
Specifically, we show that the known concentration of charged defects in 
typical oxides yields a high probability that unintentional levels will be 
present, and that one or more electrons in the impurity-induced level are 
likely to be coupled to a lithographically defined dot with sufficient 
strength to lift the spin blockade. 
We also report the
experimental observation of a magnetic field-independent chemical potential along one charge transition in one device 
studied. 
\textcolor{black}{While other mechanisms such as coupling to nuclear spins and spin-flip co-tunneling can lead to the lifting of spin blockade, these other mechanisms do not provide a natural explanation of a field-independent charging transition.}
In addition, our calculations show that modifying the device geometry to 
increase screening of charges in the oxide layer (for instance, by placing 
metal gates directly above the quantum dots, as in Refs.~\onlinecite{Veldhorst:2014eq, 
Veldhorst:2015je,Zajac:2016fh}) reduces the likelihood that impurity-induced levels affect these devices for a given density of defects in the oxide.

The paper is organized as follows.  Section~\ref{sec:methods} presents both
the experimental methods (in Subsection~\ref{subsec:experimental_methods}) and
the theoretical methods (in Subsection~\ref{subsec:theoretical_methods}).
Section~\ref{sec:results} presents the main results.
Subsection~\ref{subsec:experimental_results}
presents the measurements representative of the 9 out of 10 devices
in which large singlet-triplet splittings are observed
but there is no evidence of spin blockade.
Anomalous magnetospectroscopy data measured on one such sample are also presented,
supporting the hypothesis that there
are electrons occupying unintended
electron levels.
Subsection~\ref{subsec:theoretical_results} presents the theoretical results
that demonstrate that impurity-induced levels
provide a reasonable explanation of the experimental observations.
The results are discussed and summarized in Section~\ref{sec:discussion}.
Appendix~\ref{appendix:comsol_details} presents additional details of the
finite element calculations.
Appendix~\ref{appendix:gate_voltage_tables} provides the gate voltages used in our simulations.
Appendix~\ref{appendix:spin_blockade_observed_data} shows the results from
the one device measured that exhibited spin blockade.
Appendix~\ref{appendix:anomalous_magnetospectroscopy} presents a more
detailed discussion of the anomalous magnetospectroscopy results where
a magnetic field-independent chemical potential is observed.
Appendix~\ref{appendix:overlapping_details}
presents details of the calculations made using different gate geometries.

\section{Methods}
\label{sec:methods}
This section presents the methods used both for the experiments and for the
theoretical calculations.
\subsection{Experimental Methods}
\label{subsec:experimental_methods}
\begin{figure}[t]
	\centering
	\includegraphics[width=0.8\textwidth]{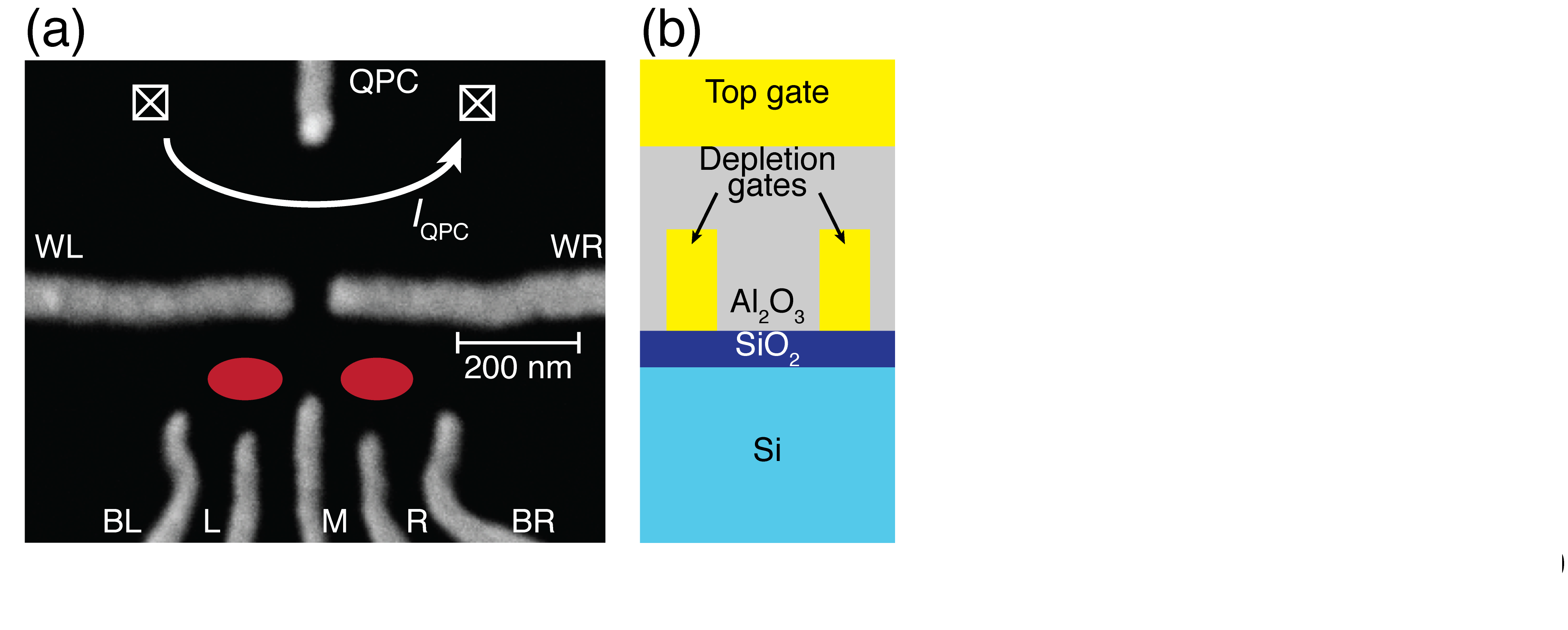}
	\vspace{-1cm}
	\caption{Experimental devices.
		(a) Scanning electron micrograph of a device with a design identical 
		to the ones measured, with the gates QPC, WL, WR, BL, L, M, R, BR labeled. 
		The approximate location of the quantum dots are indicated with red 
		ellipses.
		(b) Side-view schematic of the device structure. 
		\label{fig:setup}}
\end{figure}

	Ten nominally identical devices were fabricated and measured, each with a 
	\SI{20}{\nano\metre} layer of SiO$_2$.
	The Ti/Au electrostatic gates were created using electron-beam lithography, 
	evaporation, and liftoff techniques.
	A conformal layer of aluminum oxide was then applied,
	followed by deposition of a global top gate.
A scanning electron micrograph (SEM) of the essential part of a device, similar 
to those measured, is shown in Fig.\,\ref{fig:setup}(a). 
In this device architecture, the global top gate is used to accumulate electrons at the 
Si/SiO$_2$ interface, and a double quantum dot (DQD) system is defined by applying appropriate voltages to seven of the
confinement gates. 
A schematic of a device cross-section is shown in Fig.\,\ref{fig:setup}(b). 

The devices are operated and measured in a dilution refrigerator operating at a base 
temperature of approximately \SI{60}{\milli\kelvin}. 
The dots are characterized by measuring the differential conductance through 
the quantum point contact (QPC); peaks in the differential conductance occur
at voltages at which the occupation in a dot changes.
Pulsed gate experiments are performed in magnetic fields,
to further characterize the behavior of the samples, as described in
Section~\ref{sec:results}.

\subsection{Theoretical Methods}
\label{subsec:theoretical_methods}
Our theoretical calculations address the question of whether
impurities in the oxide layer of these devices can induce 
unintentional levels that are not easily discernible in stability
diagrams but which can cause spin blockade to be lifted.
In our calculations we assume that all conduction band valley splittings are large and include
only electrons in the lowest valley; this is consistent with experiment,
since all
devices studied here are confirmed to have valley 
splittings of at least \SI{100}{\micro\electronvolt} (see Subsection~\ref{subsec:experimental_results}) --
values that are consistent with Si-MOS devices reported in the literature~\cite{Hao:2014ea}.

For impurity-induced levels to be a reasonable explanation for the
experimental measurements, they must have large energy spacings, so that changes in occupation are not apparent in typical
stability diagrams because the occupancy of the level does not change
over the range of voltages investigated.
This requirement implies that the electron wavefunctions in the impurity-induced level must be highly localized.
At the same time, the electron in the impurity-induced
level must have reasonably strong
tunnel coupling to an electron in one of the lithographically defined dots,
so that spin blockade is lifted via the
process shown in Fig.\,\ref{fig:cartoons}(a).
In this process, 
the electron in one of the lithographic dots can go into a singlet state
in the already-occupied other dot
while conserving all the spin quantum numbers of the three-electron system 
because the spin on the lithographic dot can flip due to 
exchange with an electron in the impurity-induced level.
This behavior is closely related to processes that occur in other 
three-electron systems \cite{Shaji:2008hq, Simmons:2010ev, Koh:2011cx}, 
including the quantum dot hybrid qubit~\cite{Shi:2012kla,cao2016tunable}.

A crucial ingredient of our theoretical investigation is to estimate the 
likelihood that a given device has an impurity-induced accidental level that 
is sufficiently strongly coupled to lift spin blockade.
We note that for tunnel couplings $t$ that are small compared to the energy 
level detunings $\varepsilon$ between the gate-defined dot and the 
impurity-induced level, the rate of virtual transfer of an electron into a 
singlet state in which one dot is doubly occupied is of order $J/h$, where $J$ is
the exchange coupling between the two levels.
The exchange can be approximated as $J \approx t^2 / 
\varepsilon$, as appropriate when $|\varepsilon|$ is large compared to the 
charging energy of the dot. $\varepsilon < 0$ corresponds to a (1,1) ground 
state, where $(m,n)$ denotes $m$ electrons in the lithographic dot and $n$ 
electrons in the levels induced by the impurity. 
We compute this rate and compare it to the frequency of the square 
wave pulses used in the experiment to characterize spin blockade.
To lift spin blockade, the electron occupying the impurity level need only be 
coupled to one of the intentional dots in the system. 

The full theoretical approach, described below, involves performing simulations of the electrostatics and quantum confinement of electrons in the experimental device, both
in the absence and presence of a charged impurity.
The results of the simulations are used to determine under what conditions the
impurity-induced levels are expected to be occupied.
We also calculate the tunnel coupling
of an electron between a lithographic dot and an impurity-induced
level.

\textcolor{red}{The theoretical method is summarized as follows. We first calculate the screened electrostatic potential and 
self-consistent charge distribution using 
Thomas-Fermi simulations, while adjusting the gate voltages to obtain 
single electron occupancy in each dot, once excluding and then including an 
impurity potential.
The lowest two eigenstates of the single-particle 2D Schrödinger 
equation are then obtained for both cases.
Section~\ref{subsubsec:exchange_calculations} shows how the exchange coupling between an electron in a lithographic quantum dot and an electron in an impurity-induced state is extracted from these calculations. We perform the calculations at different possible impurity locations to be able to estimate the probability that an impurity is at a position that would lead to the lifting of spin blockade.
} 

\subsubsection{Calculation of Exchange Couplings}
\label{subsubsec:exchange_calculations}
In this subsection we present our method
for estimating $t$ and $\varepsilon$ and thus obtaining 
$J\sim t^2/\varepsilon$,
the exchange coupling $J$ between an electron in a lithographic dot and an electron in an impurity-induced level.

\begin{figure*}[t]
	  \centering
	  \includegraphics[width=1.0\textwidth]{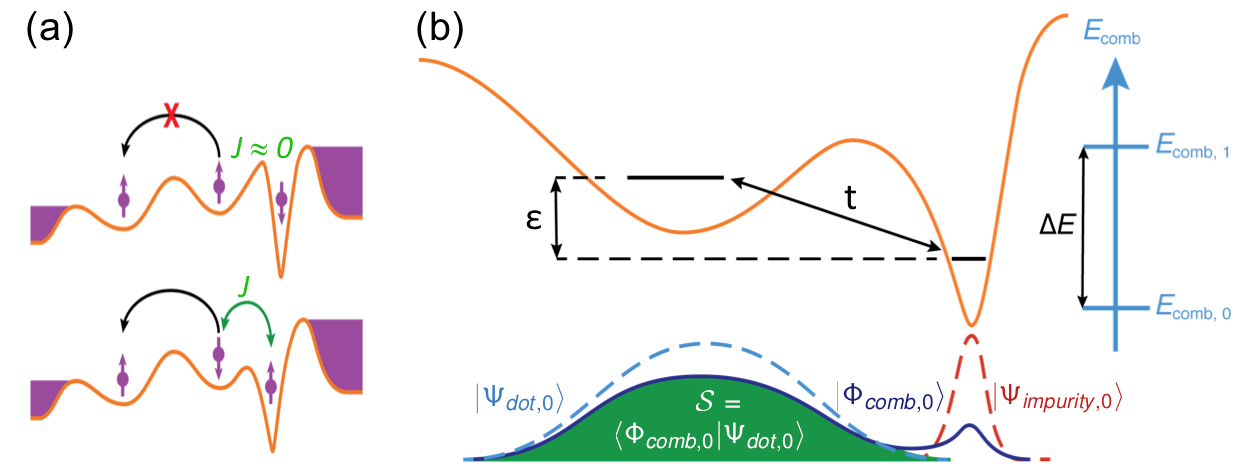}
	\caption{
	(a) Cartoons depicting the exchange process that lifts Pauli spin blockade.
	The top cartoon depicts an electron in an impurity-induced level that is not exchange-coupled and therefore does not
	lift Pauli blockade.
	The bottom cartoon illustrates how exchange enables spins to switch between 
	the lithographic dot and the impurity-induced level, allowing spin blockade 
	to be lifted. 
(b) Cartoon depicting parameters used for the calculations of the exchange coupling between the impurity level and a lithographic dot. 
	  The dashed lines indicate unperturbed basis wavefunctions $\left| 
	  \psi_{\text{dot},0} \right\rangle$ and $\left| \psi_{\text{impurity},0} 
	  \right\rangle$ used for the charge qubit Hubbard model.
	  The solid blue line depicts the ground eigenstate of the combined 
	  dot-impurity potential, indicated by the solid orange line. 
	  The overlap $\mathcal{S}$ used to calculate the exchange coupling is
	  indicated by the shaded green region.
\label{fig:cartoons}
} 
\end{figure*}

We first determine the amount of hybridization 
	between the lithographic dot ground state and the impurity-induced level 
	ground state in the combined confinement potential. 
	Let $\left|\psi_{\text{dot},0} \right\rangle$ be the unperturbed lithographic 
	dot ground state, $\left| \psi_{\text{impurity},0}\right\rangle$ be the 
	impurity level ground state, and $\left| \phi_{\text{comb},0} \right\rangle$ 
	be the hybridized ground eigenstate of the combined system.
	Assuming $\left| \phi_{\text{comb},0} \right \rangle$ can be decomposed in terms of $\left| \psi_{\text{dot},0} \right\rangle$ and $\left| 
	\psi_{\text{impurity},0}\right\rangle$, we apply a Hubbard model to 
	estimate the detuning and tunnel couplings using the energy difference between the total system ground state 
	and the lithographic-dot ground state, $\Delta E = 
	E_{\text{comb},1} - E_{\text{comb},0}$ where $E_{\text{comb},i}$ is the 
	energy of the $i$-th eigenstate of the combined dot-impurity system.
	These parameters are depicted in Fig.\,\ref{fig:cartoons}(b).

	A general form for the Hamiltonian of a two level system defined by the  
	$\left\{\left| \psi_{\text{dot},0} \right \rangle, \left| 
	\psi_{\text{impurity},0} \right\rangle\right\}$ basis is given by the 
	two-by-two matrix:
	\begin{equation}
	  H = \begin{pmatrix}
	  -\varepsilon/2 & t \\
	  t & \varepsilon/2
	  \end{pmatrix}
	  \label{eqn:diagHamil}
	\end{equation}
	with $\varepsilon$ the energy detuning between the two states of the system 
	and $t$ the tunnel coupling between them. 
	The eigenenergies of this system are $E = \pm \tfrac{1}{2} \sqrt{4t^2 + 
	\varepsilon^2}$ and the general forms for real eigenvectors are
	\begin{equation}
	  \begin{split}
	  \left| \phi_{\text{comb},0} \right\rangle &= 
	  \cos{\left(\frac{\theta}{2}\right)} 
	  \left| \psi_{\text{dot},0} \right\rangle \\
	  & \quad + \sin{\left(\frac{\theta}{2}\right)} 
	  \left| \psi_{\text{impurity},0} \right\rangle
	  \end{split}
	  \label{eqn:phicomb0}
	\end{equation}
	\begin{equation}
	  \begin{split}
	  \left| \phi_{\text{comb},1} \right\rangle &= 
	  -\sin{\left(\frac{\theta}{2}\right)} 
	  \left| \psi_{\text{dot},0} \right\rangle \\
	  & \quad + \cos{\left(\frac{\theta}{2}\right)} 
	  \left| \psi_{\text{impurity},0} \right\rangle,
	  \end{split}
	  \label{eqn:phicomb1}
	\end{equation}
	with $\theta$ the mixing angle between the lithographic dot and the impurity 
	dot.
	  
	Using this model, we can compute the energy difference $\Delta E = \sqrt{4t^2 
	+ \varepsilon^2}$ and the overlap integral $\mathcal{S} = \cos{(\theta / 2)} 
	= \left\langle \psi_{\text{dot},0} \right| \left. 
	\phi_{\text{comb},0}\right\rangle$.
	Inverting Eq.\,(\ref{eqn:diagHamil}) transformed into the basis defined by 
	Eqs.\,(\ref{eqn:phicomb0})-(\ref{eqn:phicomb1}) for $t$ and $\varepsilon$ 
	using these variables yields
	\begin{equation}
	  t = \Delta E \sqrt{\mathcal{S}^2 \left(1 - \mathcal{S}^2\right)}
	\end{equation}
	and
	\begin{equation}
	  \varepsilon = \Delta E \left(1-2\mathcal{S}^2\right).
	\end{equation}
	  
	In general, it is possible that the ground state and the first excited state 
	of the total perturbed system are not well described as combinations of the 
	unperturbed ground states of the lithographic dot and the occupied impurity-induced level. 
	The most common of these scenarios is when the impurity potential is weak 
	enough that the combined system is almost identical to that of the 
	unperturbed lithographic dot.
	These ill-defined scenarios have no impact in our estimations because we only take into account impurity configurations meeting certain criteria regarding the total electron energy and exchange coupling, as discussed later in Subsec.~\ref{subsubsec:dangerous_locations}.
	
\subsubsection{Finite-Element Simulation Methods}
\label{subsubsec:Thomas-Fermi_calculations}
To determine whether typical Si-MOS devices possess large enough charge 
impurity densities to support the creation of spurious levels containing an 
electron capable of suppressing blockade, we perform numerical simulations.
In addition to a simulation with no impurities present,
we perform a series of calculations in which
a singly-charged impurity is introduced near the active region of
the device \textcolor{black}{within the oxide layer above the interface (see 
Fig.\,\ref{fig:Comsol_methods}(a))}.
Note that the simulations do not include an additional uniform oxide charge density, since it 
contributes only an overall shift to the device potentials.
\textcolor{black}{For each impurity location, we calculate the electron charge density and} then estimate the exchange 
coupling that would exist between the lithographically-defined quantum dot and 
the induced spurious level. 

We solve for the screened electrostatic potential and 
self-consistent charge distribution of a two-dimensional electron gas (2DEG) 
located at the Si/SiO$_2$ interface using the Thomas-Fermi approximation 
\cite{Stopa:1996kr}, which is described in Appendix~\ref{appendix:comsol_details}. 
The simulations are performed within COMSOL Multiphysics~\cite{comsol}, a 
finite element simulation suite.
The simulations are repeated, while adjusting the gate voltages, until we obtain 
device tunings with approximately one electron in each dot.
\textcolor{red}{Such calculational methods been used
successfully to model nanodevices; see, e.g., Refs.~\onlinecite{Shaji:2008hq,zhao2019single}}.
Figure~\ref{fig:Comsol_methods}(b) depicts the Thomas-Fermi electron density for a 
typical double-dot tuning including an impurity.

Exploiting the reflection symmetry of the device, we now focus on just the right 
quantum dot, where we calculate the two-dimensional (2D) electrostatic potential experienced by a single electron, in the plane of the 2DEG, $V_{\text{dot}}$.
We introduce this confinement potential in a 2D Schr\"{o}dinger equation for the right dot,
\begin{equation}
\left(-\tfrac{\hbar^2}{2 m^*_e} \nabla^2 + V_{\text{dot}} \right) 
\left| \psi_{\text{dot},i} \right\rangle 
= E_{\text{dot},i} \left| \psi_{\text{dot},i}\right\rangle,
\label{eqn:dot2d}
\end{equation}
and solve to obtain the lowest two orbital eigenstates, $i = 0, 1$.
Here, $m^*_e$ is the transverse effective mass of a $z$-valley in silicon. We 
further adjust the gate voltages such that the orbital energy splitting, 
$E_{\text{dot},1} - E_{\text{dot},0}$, takes the experimentally reasonable value 
\SI[mode=text]{0.5}{\milli\electronvolt}, while still satisfying the requirement of having one electron in each dot.
The final gate voltages obtained through this procedure are listed in Appendix~\ref{appendix:gate_voltage_tables}.
We also solve the Schr\"odinger equation for an electrostatic potential $V_\text{comb}(x,y)$ that includes a single 
point impurity with charge $\pm |e|$, in addition to the potentials from lithographically defined gates:
\begin{equation}
\left(-\tfrac{\hbar^2}{2 m^*_e}\nabla^2 + V_{\text{comb}}\right) 
\left| \phi_{\text{comb},i} \right\rangle
= E_{\text{comb}, i} \left| \phi_{\text{comb},i} \right\rangle .
\end{equation}
Here, $\left|\phi_{\text{comb}, 0}\right\rangle$ and
$\left|\phi_{\text{comb}, 1}\right\rangle$ are the single-electron ground and 
excited states respectively of the combined dot-impurity system.
\textcolor{red}{Both $V_{dot}$ and $V_{comb}$ are computed in the Thomas-Fermi approximation for a realistic gate geometry, which simultaneously accounts for the image charges associated with the electrons in the lithographic dots, the 2DEG, the impurity potentials, and the nonlinear redistribution of charge density in the 2DEG, in response to changes in gate voltages and the position of the impurity.}

Finally, we use the methods of Sec.~\ref{subsubsec:exchange_calculations} to
calculate the exchange coupling between an electron in a 
lithographically defined dot and in the impurity-induced level.
In the current section, we obtain the inputs to this theory, including
the orbital energy spacing, $\Delta E = E_{\text{comb},1} - E_{\text{comb},0}$,
and the overlap integral between 
the combined system ground state and the dot ground state (in the 
absence of an impurity), 
$\mathcal{S} =
\left\langle \phi_{\text{comb}, 0} \right|
\left. \psi_{\text{dot},0} \right\rangle$.

\subsubsection{Calculation of ``Dangerous" Impurity Locations}
\label{subsubsec:dangerous_locations}

\begin{figure}[t]
	\centering
	\includegraphics[width=8cm]{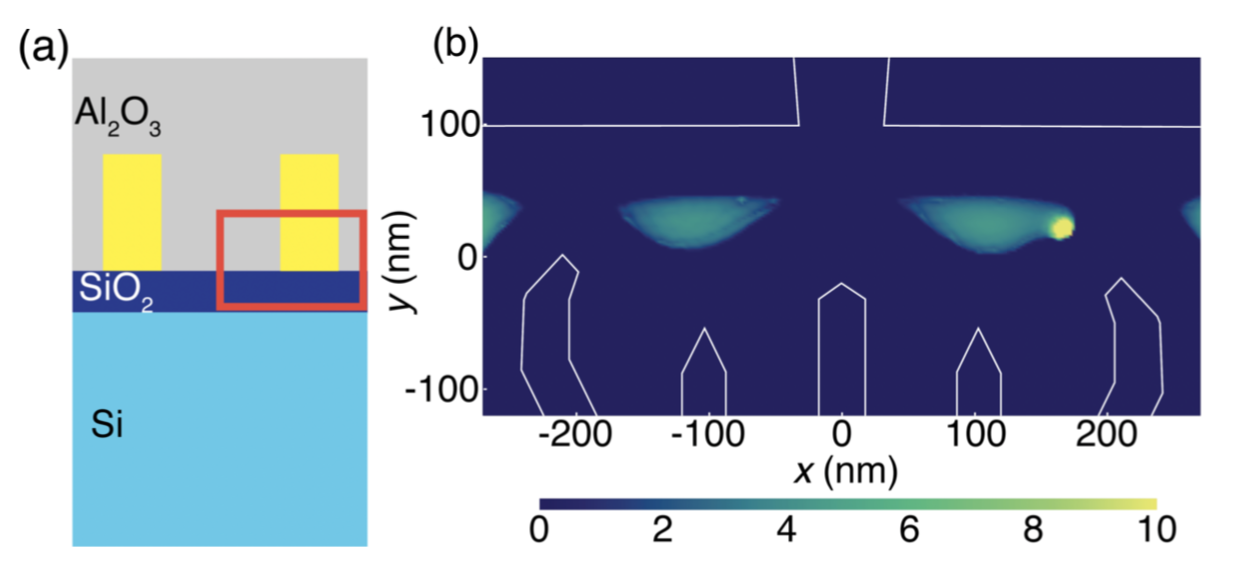}
	\caption{Method for examining locations at which a positively charged defect induces a 
		strongly coupled impurity level in the experimental devices.
		(a) Cross-section schematic of device, depicting the region 
		of oxide where charged impurities were placed (red box).
		(b) Electron density within a two-dimensional electron gas (2DEG)
		under the influence of an impurity potential calculated using a 
		self-consistent electrostatics model with the 2DEG charge obtained using 
		the Thomas-Fermi approximation. 
		\label{fig:Comsol_methods}}
\end{figure}

To map out ``dangerous'' regions, i.e., the possible locations at which impurities result in levels
that lead to the lifting of spin blockade,
we perform the calculations described in 
Secs.~\ref{subsubsec:exchange_calculations} and \ref{subsubsec:Thomas-Fermi_calculations} on
a grid of 5600 
possible impurity locations throughout the active region of the device, above 
the 2DEG, amounting to a box of dimension $\SI{250}{\nano\metre} \times 
\SI{170}{\nano\metre} \times \SI{45}{\nano\metre}$ \textcolor{black}{(see 
Figure~\ref{fig:Comsol_methods}(a))}, and we sort each location based on whether 
or not an impurity at that location would lead to the spin blockade lifting 
effects detailed above.
The method used to estimate $J$ is only accurate when the electron in 
the spurious level is strongly bound, with a large energy level splitting to 
the excited state.
Moreover, a weakly bound spurious level would be apparent
in the experimentally measured stability diagram.
Since such cases can be identified and corrected experimentally, by retuning the device, we include as one of our requirements for a ``dangerous" impurity that $\Delta E > \SI{1}{\milli\electronvolt}$. 
Additionally, since the square pulse 
frequencies in our experiments are approximately \SI{10}{\mega\hertz}, 
we require a ``dangerous" impurity to act faster than this, corresponding to a lower limit on $J$ of \SI{10}{\mega\hertz} 
or \SI{4.e-2}{\micro\electronvolt}.
We then map out the impurity locations within the testing box shown in Figure~\ref{fig:Comsol_methods}(a) that 
meet these two requirements.
In this way, we determine the lower limit of the impurity density that causes lifting of spin
blockade with probability greater than \SI{50}{\percent}. 

	\begin{figure*}[ht]
	  \centering
	  \includegraphics[width=1.0\textwidth]{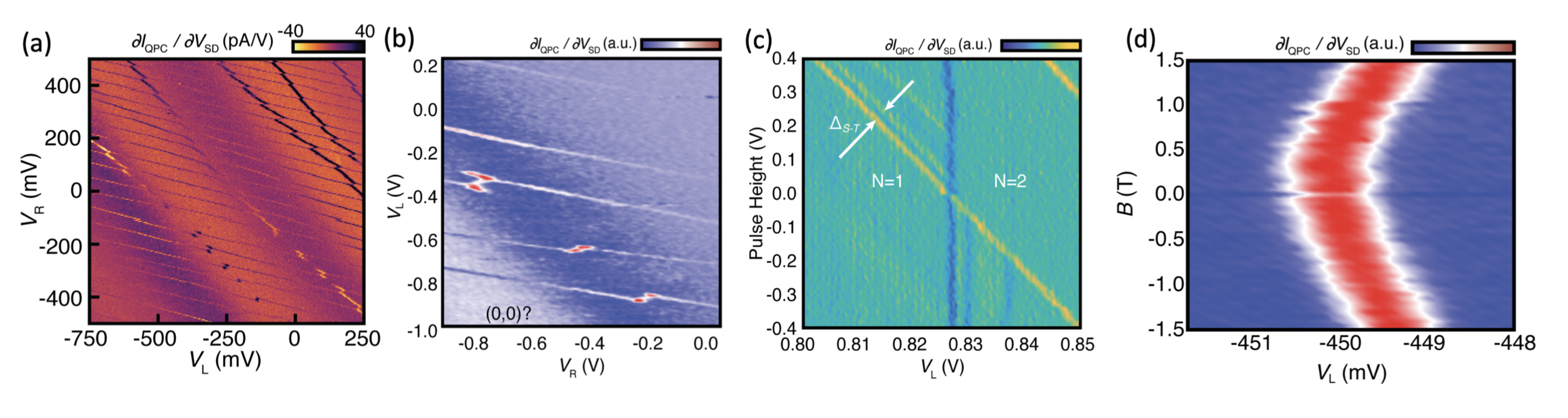}
	  \caption{Experimental measurements.
	  (a) Representative stability diagram (differential conductance through the
		quantum point contact (QPC) as a function of gate voltages $V_\text{L}$ 
		and $V_\text{R}$) for devices being considered, showing typical double dot 
		behavior.
		(b) Representative stability diagram
		showing depletion of the double dot device to low electron occupations. \textcolor{red}{The (0,0)? indicates that the occupations are consistent with being zero, but the presence of a filled shell cannot be ruled out definitively.}
	  (c) \textcolor{red}{Representative differential conductance through the QPC measured as a function 
    of the baseline voltage applied to gate L, $V_\mathrm{L}$, when a square wave pulse is added
    to the dc baseline applied to gate L.
	  This excited state spectroscopy is used to measure the 
    singlet-triplet splitting in a dot, which is extracted from the difference in voltage of the lines indicated by the two arrows.}
    For these measurements, $V_{\mathrm{L}}$ is attenuated by a factor of 10, 
    and the pulse height is attenuated by a factor of 33.
    (d) A representative magnetospectroscopy
	  plot demonstrating Zeeman splitting of an effectively single-electron 
	  quantum dot via differential conductance through the quantum point contact 
	  (QPC) as a function of the applied magnetic field $B$ and the voltage on 
	  gate L, $V_\mathrm{L}$.
	  The gate lever arms were determined using the relation that the Zeeman 
	  energy is equivalent to \SI{0.12}{\milli\electronvolt\per\tesla}.
	  Adding a second electron to the quantum dot becomes less energetically
	  favorable as the magnetic field increases, over the entire range
	  measured, providing strong evidence
	  that the singlet-triplet splitting in the dot is large.
	  \label{fig:experimental_data}}
	\end{figure*}

\section{Results}
\label{sec:results}
In this section we discuss the experimental and theoretical results.

\subsection{Experimental Results}
\label{subsec:experimental_results}

Figure~\ref{fig:experimental_data}(a) shows a stability diagram~\cite{vanderWiel:2002gra} 
in the few-electron regime \textcolor{black}{with source-drain bias of 
$V_{\mathrm{SD}} = \SI{0.5}{\milli\volt}$}, demonstrating the characteristic 
features of a double quantum dot: charging lines with two different slopes,
that occur at voltages at which an electron is added to one of the dots.

	For each of the ten devices examined, the voltages were tuned to create a 
	double dot configuration and then depleted to low electron occupations, as 
	shown in Fig.\,\ref{fig:experimental_data}(b). 
	Efforts were made to fully deplete the quantum dots; 
	for example, all 
experiments testing for Pauli blockade were performed over multiple  
anti-crossings \cite{Jones:2018vc}.
\textcolor{red}{Using the voltage differences between the transition lines together with the relevant lever arms to estimate the dot capacitances~\cite{hanson2007spins}, we find values consistent with single electron occupation.
While we cannot definitely rule out the possibility of closed shells of electrons at the nominal (0,0) occupation, the presence of closed shells does not preclude the measurement of Pauli spin blockade, as has been demonstrated in Refs.~\onlinecite{kodera2009pauli,HarveyCollard:2017ic,chen2017spin,petit2019universal,yang2019silicon,leon2020coherent}.}
	
	\textcolor{red}{Excited state spectroscopy measurements \cite{huebl2010electron,Xiao:2010cx,Maune:2012iu},
	shown in Fig.\,\ref{fig:experimental_data}(c), were carried out to read the 
	singlet-triplet splitting in each dot of every device.
	In these experiments, a square pulse 
  with frequency approximately equal to the tunneling rate (a few hundred 
	\si{\hertz}) was applied in combination with an average (DC) voltage shift on gate L, to 
	populate the excited states of the quantum dot.}
	The energy difference between the ground and first excited states, corresponding to the 
	singlet-triplet splitting $\Delta_{S-T}$, was consistently found to be between 100 and 
	\SI{300}{\micro\electronvolt},
	using lever arms determined in the magnetospectroscopy experiments, as described below.
(A lever arm describes the conversion factor between a gate voltage and a relevant dot energy.)
Despite these large singlet-triplet splittings, spin blockade was not observed in nine out of ten devices.
Results from the tenth device, which did exhibit spin blockade, are presented in Appendix~\ref{appendix:spin_blockade_observed_data}.

Magnetospectroscopy experiments were conducted with the magnetic field $B$ 
oriented parallel to the dot axis \textcolor{black}{on three devices}.
Figure~\ref{fig:experimental_data}(d) shows magnetospectroscopy data used
to determine gate lever arms in one device;
these data also demonstrate that the singlet-triplet splitting in the device
is substantial, because increasing the magnetic field makes it energetically less favorable to add a new electron to form a singlet state; this allows us to unambiguously identify the ground two-electron state as a singlet~\cite{hanson2007spins}.

\begin{figure}[ht]
	\centering
\includegraphics[width=\columnwidth]{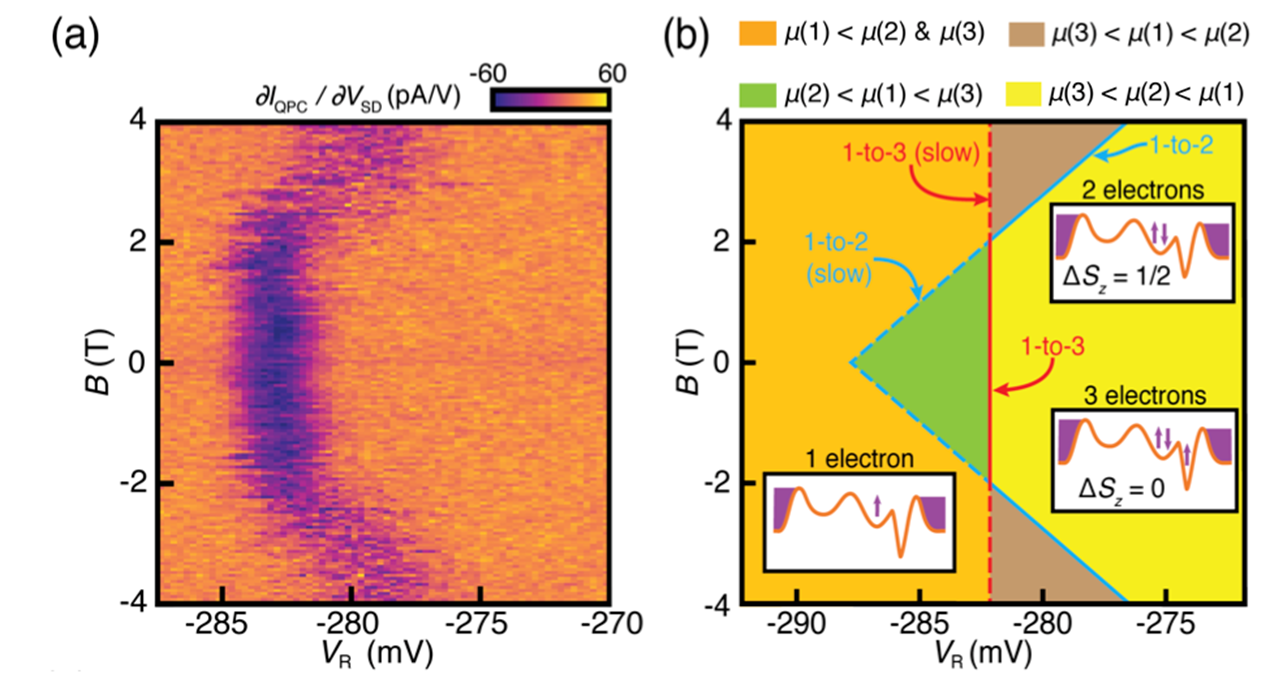}
	\caption{Additional evidence for the presence of unintentional levels. 
	(a) Magnetospectroscopy data showing the differential conductance through 
	the quantum point contact as a function of magnetic field, $B$, and the 
	voltage, $V_R$, on gate R.
	The vertical field-independent portion of the transition line is unexpected 
	when a single electron is added.
	The change in behavior is evidence of  1-to-2 and 1-to-3 electron 
	transitions. 
	(b) Cartoon of a scenario consistent with the structure observed in (a).  
	$\mu(N)$ denotes the energy to load $N$ electrons into the system 
	consisting of a lithographically defined dot and an impurity-induced 
	level.
	Regions with different energy orderings are color-coded, as shown. 
	Slow tunnel rates (dashed lines) make certain transitions invisible, as  
	discussed in the main text.
	\label{fig:modelproof}}
\end{figure}

\textcolor{red}{While typical magnetospectroscopy data are shown in Fig.~\ref{fig:experimental_data}(d),
one region of the stability diagram of one sample exhibited anomalous behavior,
as shown in Fig.\,\ref{fig:modelproof}(a), where the 
$B$-dependent line segments normally correspond to a charging transition from 
1-to-2 electrons (see Fig\,\ref{fig:modelproof}(b)). 
In this case, the transition has an unexpected segment that is independent of 
magnetic field. 
Including the presence of an occupied impurity-induced level 
provides a natural explanation for this behavior: the vertical line would 
correspond to a 1-to-3 electron transition with $\Delta S_z=0$. 
This simultaneous change of occupation of a lithographic dot and impurity level is rare because the
voltage range in a typical experimental sweep is not sufficient to change the occupation of the impurity level.}

Figure \ref{fig:modelproof}(b) shows possible energy orderings of states with 
1, 2, and 3 electrons, as pictured in the insets, for a system with one dot 
and an impurity level.
The figure contains shaded regions that correspond to different energy orderings, which
are discussed in detail in Appendix~\ref{appendix:anomalous_magnetospectroscopy}.
Charging transitions to states with more electrons occur as the gate 
voltage $V_R$ becomes less negative.
In magnetospectroscopy experiments, we would normally expect the transition to 
the three-electron configuration to occur on the far-right-hand side of the 
diagram, although its exact location depends on the relative sizes of the 
Zeeman, orbital and charging energy terms.
In particular, if the quantum dot confinement potential is weak and an extra level is present, then the charging energy for adding two 
electrons can be small, bringing the 1-to-3 electron transition into the 
measurement window, as indicated in Fig.\,\ref{fig:modelproof}(b).
While we do not measure the individual energy terms here, 
Appendix~\ref{appendix:anomalous_magnetospectroscopy}
demonstrates that reasonable values of the system parameters are consistent with the transitions 
observed in Fig.\,\ref{fig:modelproof}(a).
[The parameters listed in Appendix~\ref{appendix:anomalous_magnetospectroscopy} are used to plot
Fig.\,\ref{fig:modelproof}(b).]
The pattern of the transition line is reproduced if one assumes that the 1-to-2 
transition rate becomes unobservably slow when the voltage $V_R$ is made more 
negative and the 1-to-3 transition rate also becomes slower as the magnetic field is 
increased.
Such dependencies are not unexpected, because tunnel rates typically decrease 
as depletion gate voltages are made more negative, and a decrease in wave 
function overlap between the dot and the impurity level is expected due to 
magnetic confinement.

\subsection{Theoretical Results}
\label{subsec:theoretical_results}
\begin{figure}[t]
	\centering
	\vspace{-1.5cm}
	\includegraphics[width=\textwidth]{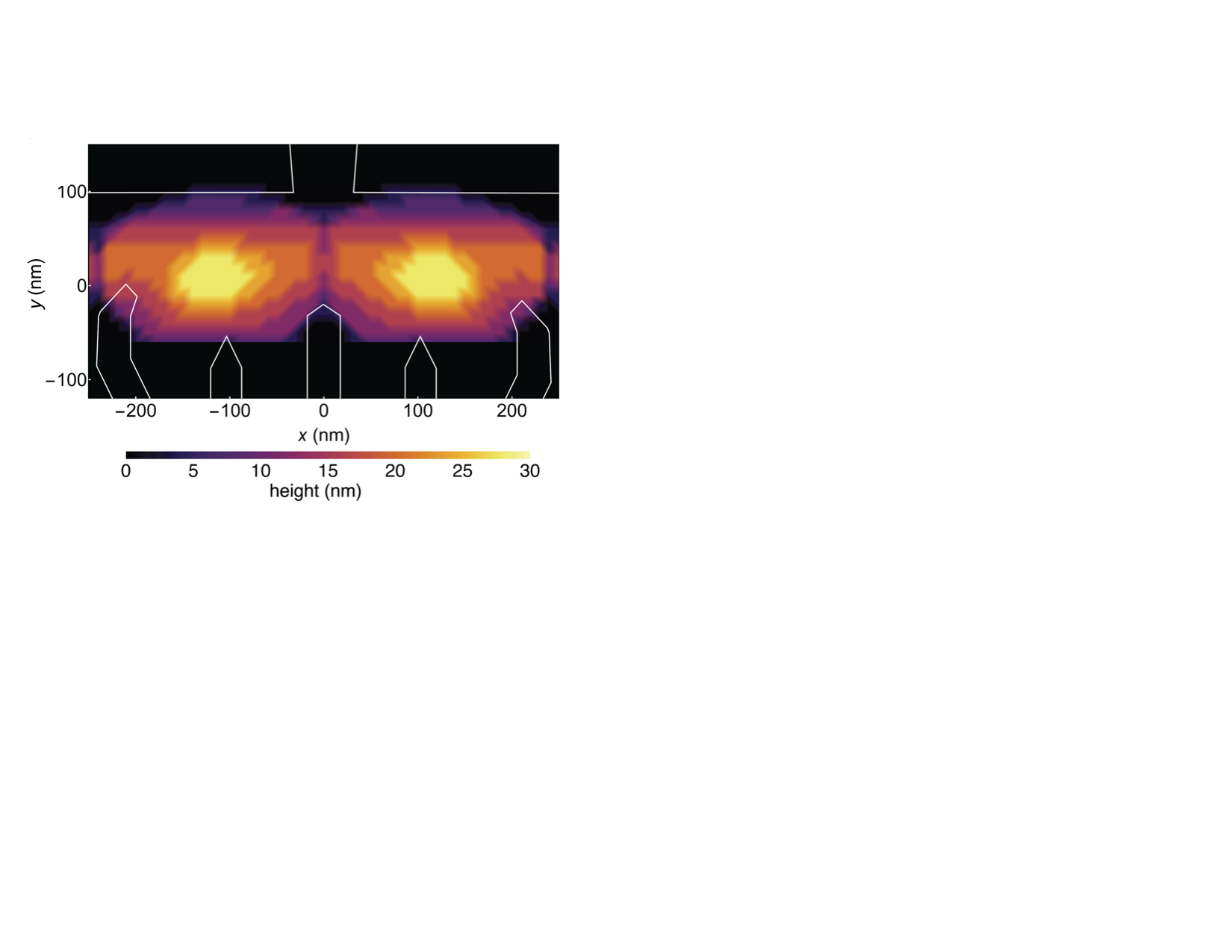}
	\vspace{-7cm}
	\caption{Characterization of the propensity for 
	a positively charged defect to induce a 
		strongly coupled impurity level that would
		lift spin blockade in the experimental devices.
The color scale corresponds to the height above the 2DEG of the dangerous regime for positively
		charged impurities.
		An impurity with charge $+e$ occurring anywhere below this height and 
		within the colored region will induce a dot capable of lifting spin 
		blockade; we consider this the dangerous region for impurities. 
		The total volume of this region is approximately  
		\SI{1.2e6}{\nano\metre\cubed} corresponding to a minimum impurity density 
		of \SI{8.6e14}{\per\centi\metre\cubed} above which spin blockade is 
		expected to be lifted. 
		\label{fig:danger_map}}
\end{figure}

Our simulations reveal some trends of interest. 
First, negative charges rarely induce unintentional levels, except when the impurity is within $\SI{5}{\nano\metre}$ of the 
2DEG and close to the center of a lithographically defined dot.
In contrast, positively charged impurities in many
different locations in the oxide induce impurity
levels that can lift spin blockade, as shown in Fig.\,\ref{fig:danger_map}\textcolor{black}.
There is a large region directly over the 2DEG where placement of a positively 
charged impurity induces an occupied impurity level with a tightly bound state 
that also has a strong enough exchange coupling that spin exchange occurs on a 
timescale less than \SI{1}{\micro\second} with a nearby gate-defined dot.
If we assume one electron within this volume, we find that a uniform impurity 
density of \SI{8.6e14}{\per\centi\metre\cubed} causes lifting of spin blockade 
with high probability.
The Si/SiO$_2$ interface is a region of concern for MOS-based spin qubits and 
if we consider the sampled locations nearest to this interface, we find a 
uniform surface impurity density of \SI{1.1e9}{\per\centi\metre\squared} would 
likely result in an impurity occurring within the spin-blockade lifting area.
This density is on the low end of the expected impurity densities for these 
devices \cite{Sakaki:1970ii} -- approximately \SI{e10}{\per\centi\metre\squared} 
in high quality thermal oxides \cite{Kim:2017hh}.
\textcolor{black}{Comparing the calculated threshold and the expected value 
suggests a high likelihood of spin blockade being absent from these devices.}

We stress that the analysis presented here focuses on fixed oxide charges, which are only one particular class of Si/SiO$_2$ interface traps. A direct comparison between this charge density threshold and typical values of densities of interface traps (as measured in cm$^{-2}$ eV$^{-1}$ by CV methods, for instance) may not be direct, requiring further identification of the physical mechanism behind the different traps. Other trapping mechanisms, such as fast interface states and chemical bonding faults right at the Si/SiO$_2$ interface are also likely to affect the spin state of qubits, but are out of the scope of the current work.

Modifying the gate geometry to increase the screening of the oxide layer reduces 
the likelihood that spin-blockade lifting occurs via this mechanism.
To demonstrate this, we examine two modifications to the gate structure: 
moving the global top gate from \SI{100}{\nano\metre} above the 2DEG to 
\SI{50}{\nano\metre}, and altering the gate layout to an overlapping gate 
design similar in layout to that used in a Si-MOS device\cite{Veldhorst:2014eq, 
Veldhorst:2015je}.
The details of the gate layout we consider are adapted from a device fabricated on 
Si/SiGe \cite{Zajac:2016fh}. 
(Further details on 
gate geometries studied here are provided in Appendices~\ref{appendix:gate_voltage_tables} and \ref{appendix:overlapping_details}.)
Generally, we find that using an overlapping gate design has the largest 
impact on our results, due to the compact coverage of metallic gates directly above the 2DEG 
and the overall tighter confinement of the lithographically defined dots 
compared to the original device considered.
The likelihood that spin blockade is lifted is found to decrease, 
as quantified by the impurity density threshold increasing by a factor 
of eighteen, while the interface impurity density threshold increases by 
a factor of three. 
\textcolor{black}{Increasing accumulation gate voltages could also 
increase the likelihood of observing spin blockade because this increases 
confinement within a dot, which in turn reduces the wavefunction overlap between 
the lithographic dot and an occupied impurity-induced level.}
Another route for increasing the device yield is to improve the quality of the 
oxide, particularly at the Si/SiO$_{2}$ interface, since this reduces the 
number of charges in close proximity to the lithographically-defined quantum 
dots.
Finally, moving the active region of the device further away from the 
impurities, by using a Si/SiGe heterostructure for example, also reduces the 
likelihood of forming dangerous impurity dots.

\section{Discussion}
\label{sec:discussion}
Our measurements and calculations indicate that failure to observe
spin blockade in a substantial fraction of Si-MOS 
double quantum dot devices with large singlet-triplet splittings
could arise because of additional energy levels induced by
impurities in the oxide of the devices.
We show that in the samples we investigated,
there is a reasonable probability that
unintentional levels produced by trapped positive charges in the 
oxide layer have large enough binding energies that they would
not typically be apparent in charge stability diagrams, and yet their
exchange coupling to one of the lithographically defined dots is
large enough to cause suppression of
spin blockade.
Typical densities of defects in these devices are consistent with the 
observations.

This problem can be mitigated not only by altering fabrication methods
to reduce the number of charged defects in the device oxide.
Our calculations indicate that employing device designs in which metal gates 
are positioned directly over the dots, which enhances the screening of the 
impurity levels, also can mitigate the problem substantially.

\begin{acknowledgments}
	The authors thank M.\ Eriksson, A.\ Frees, and J.\ S.\ Kim for useful comments and discussions.
	This work was supported by ARO (W911NF-12-1-0607, W911NF-17-1-0274, 
	W911NF-14-1-0346, W911NF-17-1-0242, W911NF-12-1-0609, W911NF-17-1-0257), 
	NSF (OISE-1132804), the Department of Defense under 
	Contract No.\ H98230-15-C0453, MEIC (Spain) FIS2012-33521 
	and FIS2015-64654-P, CSIC (Spain) Research Platform PTI-001, and CNPq (Brazil) 309861/2015-2 and 304869/2014-7. 
	The authors acknowledge support from the Vannevar Bush Faculty Fellowship 
	program sponsored by the Basic Research Office of the Assistant Secretary 
	of Defense for Research and Engineering and funded by the Office of Naval 
	Research through grant N00014-15-1-0029.
	The views and conclusions contained in this document are those of the authors and should not be interpreted as representing the official policies, either expressed or implied, of the U.S. Government. The U.S. Government is authorized to reproduce and distribute reprints for Government purposes notwithstanding any copyright notation herein.
\end{acknowledgments}

\appendix 

\section{Details of Thomas-Fermi Simulations}
\label{appendix:comsol_details}
In this appendix, we present additional details of the Thomas-Fermi simulations.

As described in the main text, we performed simulations of either two electrons in a lithographically defined double quantum dot, 
	or three electrons in a double dot system, with an additional impurity-induced 
	level.
	To begin, we treat the double dot two-electron system semi-classically as 
	described below.
	  
	The three-dimensional (3D) finite element simulations are conducted on a 
	\SI{2}{\micro\metre} by \SI{2}{\micro\metre} section of the active region of 
	the device. 
	The device stack consists of \SI{200}{\nano\metre} of silicon, 
	\SI{20}{\nano\metre} of SiO$_2$ and \SI{100}{\nano\metre} of Al$_2$O$_3$. The 
	boundary conditions used in all 3D models are as follows. The bottom surface of the 
	device is set to $V=0$; for all other boundaries, we apply the conditions $\mathbf{D\cdot n} = 0$, where 
	$\mathbf{D}$ is the electric displacement field and $\bf{n}$ the unit normal at 
	the surface.
	In all cases, the simulation cell boundary was chosen large enough that the exact position of the boundaries had no effect on our results. 
	  
	We first determine a set of gate voltages, as listed in Appendix~\ref{appendix:gate_voltage_tables}, which cause approximately one electron to accumulate in each dot.
	(Since this is a semi-classical calculation, electron quantization must be enforced by hand.)
	The 2DEG for this system is assumed to form at the Si/SiO$_2$ 
	interface and the charge density is calculated self-consistently by applying 
	the Thomas-Fermi approximation \cite{Stopa:1996kr} to model the areal 
	charge density $\sigma$, using the following definition:
	\begin{equation}
	  \sigma_{\text{TF}} = \left\{ \begin{array}{cc} -\frac{e m^{*}_{e}}{\pi \hbar^2} (eV), & \quad V > 0 \\
	  0, & \rm(otherwise) \end{array} \right. ,
	  \label{eqn:tf}
	\end{equation}
	where $m^{*}_{e}$ is the 2D transverse effective mass of a $z$-valley electron in 
	silicon and $V(x,y)$ is the spatially varying electrostatic potential at the position of the 2DEG.
	Here, we assume the valley splitting is large enough that we only need to consider one 
valley state.
	In Eq.\,(\ref{eqn:tf}), we define the electron Fermi level to occur at $V = 0$.

\begin{figure}[t]
	  \centering
	  \includegraphics[width=8.6cm]{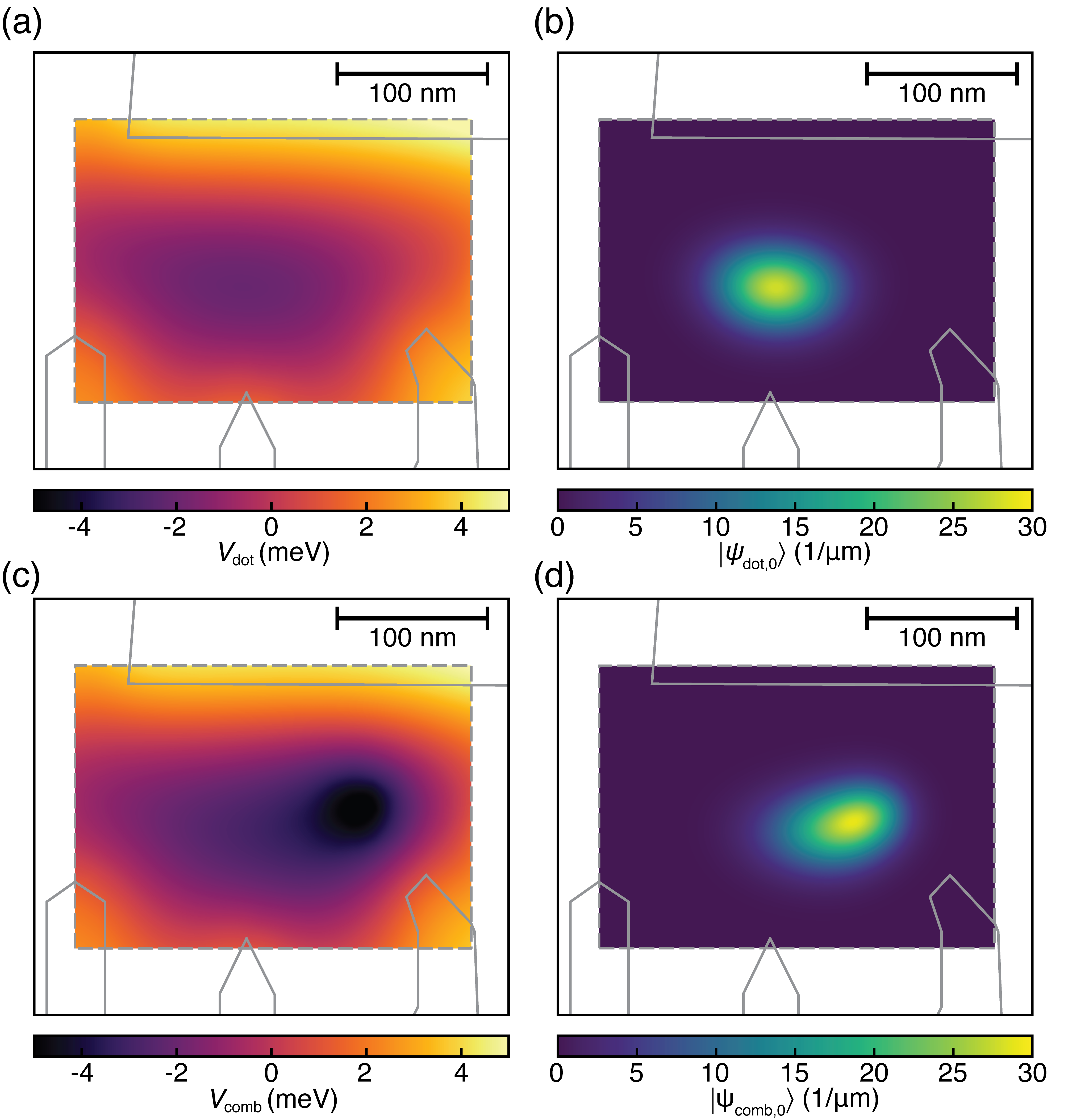}
	  \caption{Potential landscape and wavefunctions used in the simulations. 
	  Solid gray lines indicate the gate structure, and the dashed gray line 
	  demarcates the simulation regime used for the 2D Schr\"odinger simulations.
	  (a) Potential landscape for the right quantum dot, $V_\text{dot}$
	  (b) Ground eigenstate of the right quantum dot, $\left| 
	  \psi_{\text{dot},0}\right\rangle$.
	  (c) Representative potential landscape of right dot plus an impurity, 
	  $V_{\text{comb}}$.
	  (d) Surface plot showing the ground eigenstate of $V_{\text{comb}}$, 
	  $\left| \phi_{\text{comb}, 0}\right\rangle$.
	  \label{fig:calcdetail}}
	\end{figure}

As described in the main text, the results of the Thomas-Fermi electrostatics simulations are evaluated in the plane of the 2DEG, and input into two-dimensional (2D) Schr\"odinger equations. The full procedure is repeated, with and without introducing an impurity into the device, and considering a 3D grid of impurity locations.
Some typical electrostatic and quantum simulation results are shown in Fig.\,\ref{fig:calcdetail}, for the right-hand dot. 
	  
	\section{Gate Voltage Tables}
	\label{appendix:gate_voltage_tables}
	In this appendix, we list the gate voltages used in our simulations, which were chosen to accumulate one electron in each dot, while maintaining an orbital excitation energy in the right dot corresponding to 0.5~meV.
	The three tables provide the voltages used for the gate geometry used in the experiment,
	for a geometry similar to that used in the experiment except that the global top gate is moved 50 nm closer to the 2DEG, and for an overlapping gate design similar to the one used in Ref.~\onlinecite{Zajac:2016fh}.
	
	\begin{table}[h]
	  \centering
	  \begin{tabularx}{0.4\textwidth}{X X}
	  	\hline
	  	\hline
	  	\textbf{Gate} & \textbf{Voltage} (\si{\milli\volt}) \\
	  	\hline
	  	$V_{\text{Top}}$      & 98    \\
	  	$V_\text{QPC}$        & -250  \\
	  	$V_\text{WL}$         & -14.4 \\
	  	$V_\text{WR}$         & -14.4 \\
	  	$V_\text{BL}$         & -8    \\
	  	$V_\text{L}$          & -10   \\
	  	$V_\text{M}$          & -10.4 \\
	  	$V_\text{R}$          & -10   \\
	  	$V_\text{BR}$         & -8    \\
	  	\hline
	  	\hline
	  \end{tabularx}
	  \caption{Gate voltages that yield single electron occupation with $\approx 
	  \SI{0.5}{\milli\electronvolt}$ orbital splitting in the quantum dot, for
	  the gate geometry used in the experiments.}
	\end{table}
	  
	\begin{table}[thb]
	  \centering
	  \begin{tabularx}{0.4\textwidth}{X X}
	  	\hline
	  	\hline
	  	\textbf{Gate} & \textbf{Voltage} (\si{\milli\volt}) \\
	  	\hline
	  	$V_{\text{Top}}$     & 45   \\
	  	$V_\text{QPC}$       & -150 \\
	  	$V_\text{WL}$        & -30  \\
	  	$V_\text{WR}$        &  -30 \\
	  	$V_\text{BL}$        & -34  \\
	  	$V_\text{L}$         & -5   \\
	  	$V_\text{M}$         & -40  \\
	  	$V_\text{R}$         & -5   \\
	  	$V_\text{BR}$        & -32  \\
	  	\hline
	  	\hline
	  \end{tabularx}
	  \caption{Gate voltages that 
	  yield single electron occupation with $\approx 
	  \SI{0.5}{\milli\electronvolt}$ orbital splitting in the quantum dots
	  for the design modification in which the top gate is moved closer
	  to the quantum dots.}
	\end{table}

	\begin{table}[hbt]
	  \centering
	  \begin{tabularx}{0.4\textwidth}{X X}
	  	\hline
	  	\hline
	  	\textbf{Gate} & \textbf{Voltage} (\si{\milli\volt}) \\
	  	\hline
	  	$V_\text{S1}$      & 0    \\
	  	$V_\text{SL}$      & -60  \\
	  	$V_\text{SR}$      & -60  \\
	  	$V_{\text{SD},1}$  & 400  \\
	  	$V_{\text{SD},2}$  & 400  \\
	  	$V_\text{BL}$      & -12  \\
	  	$V_\text{L}$       &  235 \\
	  	$V_\text{M}$       & -7   \\
	  	$V_\text{R}$       & 235  \\
	  	$V_\text{BR}$      & -12  \\
	  	\hline
	  	\hline
	  \end{tabularx}
	  \caption{Gate voltages for the overlapping gate design that yield single 
	  electron occupation with $\approx \SI{3.5}{\milli\electronvolt}$ orbital 
	  splitting in the quantum dots.}
	\end{table}
	  
\section{Experimental Evidence for Spin Blockade in One Device}
\label{appendix:spin_blockade_observed_data}
\begin{figure}[ht]
	  \centering
	  \includegraphics[width=8.5cm]{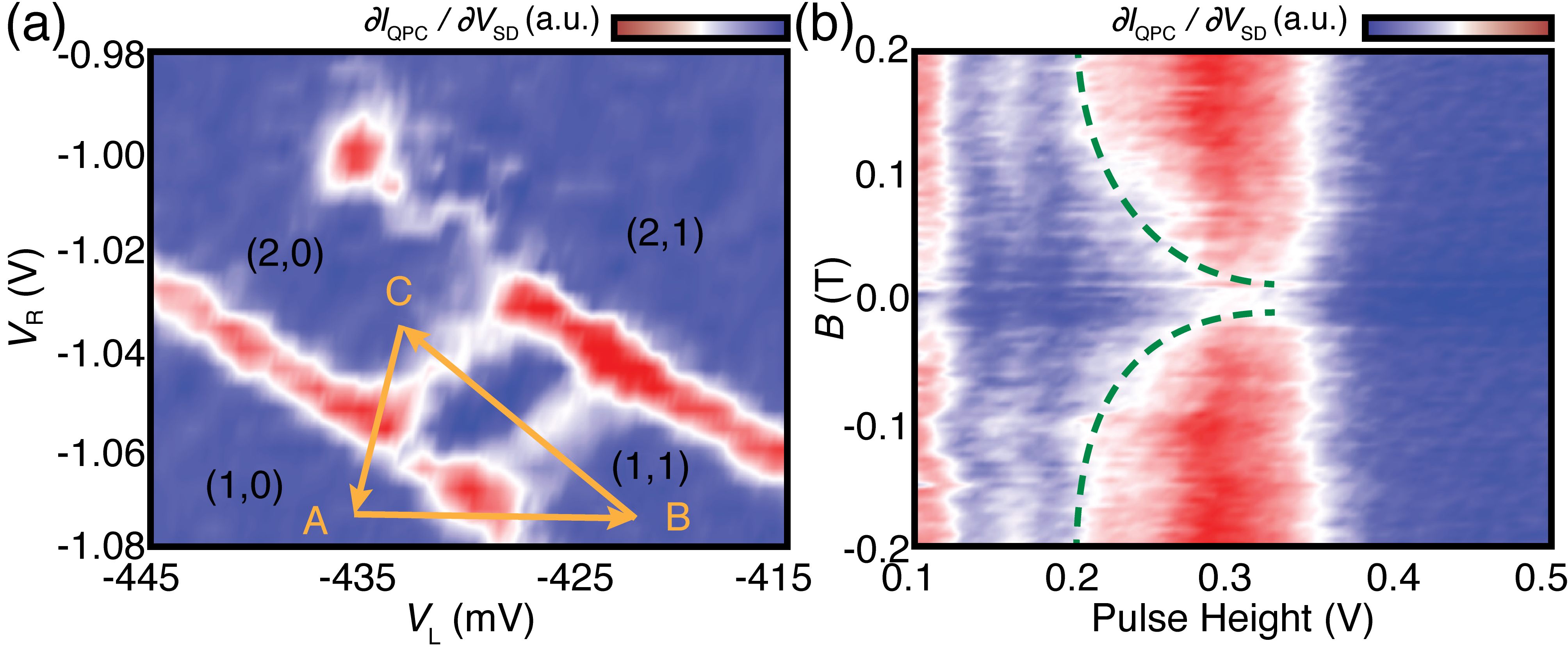}
	  \caption{Evidence of Pauli spin blockade in one device of 
	  a batch of ten.
	  (a) Search procedure for spin blockade. Shown is the differential 
	  conductance through the QPC as a function of the voltages on gates L and 
	  R, $V_\mathrm{L}$ and $V_\mathrm{R}$. The $(m,n)$ notation indicates 
	  the number of electrons (possibly over closed shells), with $m$ the 
	  number of electrons on the left dot and $n$ the number of electrons on the 
	  right dot.
	  The cycle is: (1) wait at A for \SI{500}{\nano\second}, (2) pulse to B 
	  and mix in (1,1) for \SI{1}{\micro\second}, and (3) pulse to C and measure for 
	  \SI{20}{\micro\second}. 
    This device exhibits two split transition lines, which is a signature of Pauli spin blockade.  However, the other nine devices that were measured did not.
	  (b) A spin funnel experiment also provides indication of Pauli spin 
	  blockade in this device.
	  Enhanced conduction occurs near the anti-crossing between the singlet and the
	  spin-polarized triplet at different magnetic fields (along the green dashed line,
	  which is a guide to the eye); this enhancement in the conduction requires spin
	  blockade.
	  In this experiment, the arbitrary waveform generator was gate-modulated to 
	  enhance the signal.
	  \label{fig:psb}}
	\end{figure}

A search for Pauli blockade via a three-pulse sequence~\cite{Jones:2018vc}
was conducted for ten devices. 
This appendix reports the data demonstrating the presence of Pauli spin blockade
in one device.
Fig.~\ref{fig:psb}(a) shows a portion of a stability diagram taken in the presence
of a pulse pattern consisting
of three sequences generated on an arbitrary 
	waveform generator with two channels controlling the voltage on gate L and R; 
	the details of the sequence are shown in Fig.\,\ref{fig:psb}(a).
	The splitting of the polarization line shown in Fig.\,\ref{fig:psb}(a)
	demonstrates that there are two different voltages at which an electron
	transfers between the dots, which arises because Pauli spin blockade implies
	that a portion of the time the two-electron state is a triplet, which has
	a higher energy than the singlet.
	Moreover, a spin funnel is also observed (see 
	Fig.\,\ref{fig:psb}(b)~\cite{Petta:2005kn}), which arises
	because of the change of electron transfer between the dots at the
	anticrossing between the singlet and polarized triplet, the
	visibility of which relies on spin blockade.
	  
\section{Additional Discussion of Anomalous Magnetospectroscopy Results}
\label{appendix:anomalous_magnetospectroscopy}
In this appendix we provide a detailed discussion of our interpretation of
the anomalous magnetospectroscopy data presented in Fig.~\ref{fig:modelproof}
using our model that includes an impurity-induced level.
We do this by relating the slopes of the transition lines in the stability
diagram to the energetics of the system with up to three electrons.
The intuition behind the argument is that the most straightforward way to obtain a magnetic-field-independent voltage at which 
the charge in the system changes is to have the charge occupation change by two electrons,
and
such a change can occur if there is a significant exchange interaction
between the two electrons
that are added.

We use a simple Hubbard model to calculate the chemical potential for up to 
	three electrons occupying a system consisting of a lithographic quantum dot 
	and an impurity-induced level.
	Let $\mu(N)$ be the chemical potential of $N$ electrons in the combined 
	system.
	These chemical potentials have several dominant contributions, 
  given by
	\begin{gather}
	  \mu(1) = -\alpha_D V_g - \tfrac{1}{2}g \mu_B B, \\
	  \mu(2) = E_{\text{offset},1} - 2 \alpha_\text{D} V_g, \\
	  \mu(3) = E_{\text{offset},2} - (2 \alpha_D + \alpha_I) V_g
	  - \tfrac{1}{2} g \mu_B B - J,
	\end{gather}
	where $E_{\text{offset},i}$ is the effect of Coulomb repulsion on the total 
	system due to the previous $i$ electrons, $\alpha_D$ is the lever arm of the 
	dot, $\alpha_{I}$ is the lever arm of the impurity, $J$ is the exchange 
	energy that between the dot and the impurity for the three electron state, 
	$V_{g}$ is the voltage on the gate, $g$ is the Land\'e $g$-factor for 
	silicon, and $\mu_{B}$ is the Bohr magneton. 
	We assume that the one electron state is initialized in the spin down (up) 
	state for positive (negative) applied magnetic field, the two electron state 
	corresponds to a singlet in a single dot for the range of fields studied, and 
	that the three electron state initializes in the manifold with total spin 
	$1/2$ and $z$-component spin down (up) based on positive (negative) applied 
	field. 
	Using the parameter values in Table \ref{tbl:param}, we find the energy 
	regions shown in Fig.~\ref{fig:modelproof}(b). 
	The value of $\alpha_{D}$ was set by the slope of the data in Fig.~\ref{fig:modelproof}(a).
	The remaining parameters were tuned to values that reproduce the 
	magnetospectroscopy in Fig.~\ref{fig:experimental_data}(d) of the main text.
	Visually similar behavior can be found for different parameter choices.
	As a note, for this set of parameters, we also observe a small region where 
	$\mu(2) < \mu(3) < \mu(1)$, which outside the voltage range plotted in Fig.~\ref{fig:experimental_data}(d). 
	We assume that the same slow tunnel rate that makes the 1-to-2 transition 
	invisible, as discussed in the main text, also suppresses transitions to this 
	two-electron ground state. 
	  
	\begin{table}[ht]
	  \centering
	  \begin{tabularx}{0.4\textwidth}{X X}
	  	\hline
	  	\hline
	  	\textbf{Parameter} & \textbf{Value} \\
	  	\hline
	  	$g$ & 2.1 \\
	  	$\alpha_{D}$ & \SI[per-mode=symbol]{0.035}{\milli\electronvolt \per 
	  			\milli\volt} \\
	  	$\alpha_{I}$ & \SI[per-mode=symbol]{0.1}{\milli\electronvolt \per 
	  			\milli\volt}\\
	  	$E_{\text{offset},1}$ & \SI{0.45}{\milli\electronvolt}\\
	  	$E_{\text{offset},2}$ & \SI{2.42}{\milli\electronvolt}\\
	  	$J$ & \SI{0.3}{\milli\electronvolt}\\
	  	\hline
	  	\hline
	  \end{tabularx}
	  \caption{Parameters used for calculating chemical potentials for 1, 2, and 
	  3 electrons in a lithographic dot plus impurity-induced level system,
	  used to interpret anomalous magnetospectroscopy results shown in
	  Fig.~\ref{fig:modelproof}.
	  \label{tbl:param}}
	  \end{table}

\section{Calculations for Modified Gate Designs}
\label{appendix:overlapping_details}
	  Here we investigate gate designs that differ from the one used for the samples studied in the main text. We find that 
	  changing the gate geometry can change substantially
  the size and shape of the region where placement of an impurity is likely to 
  induce a level containing a spin blockade lifting electron.
  
\begin{figure}[t]
	  \centering
	  \includegraphics[width=8.5cm]{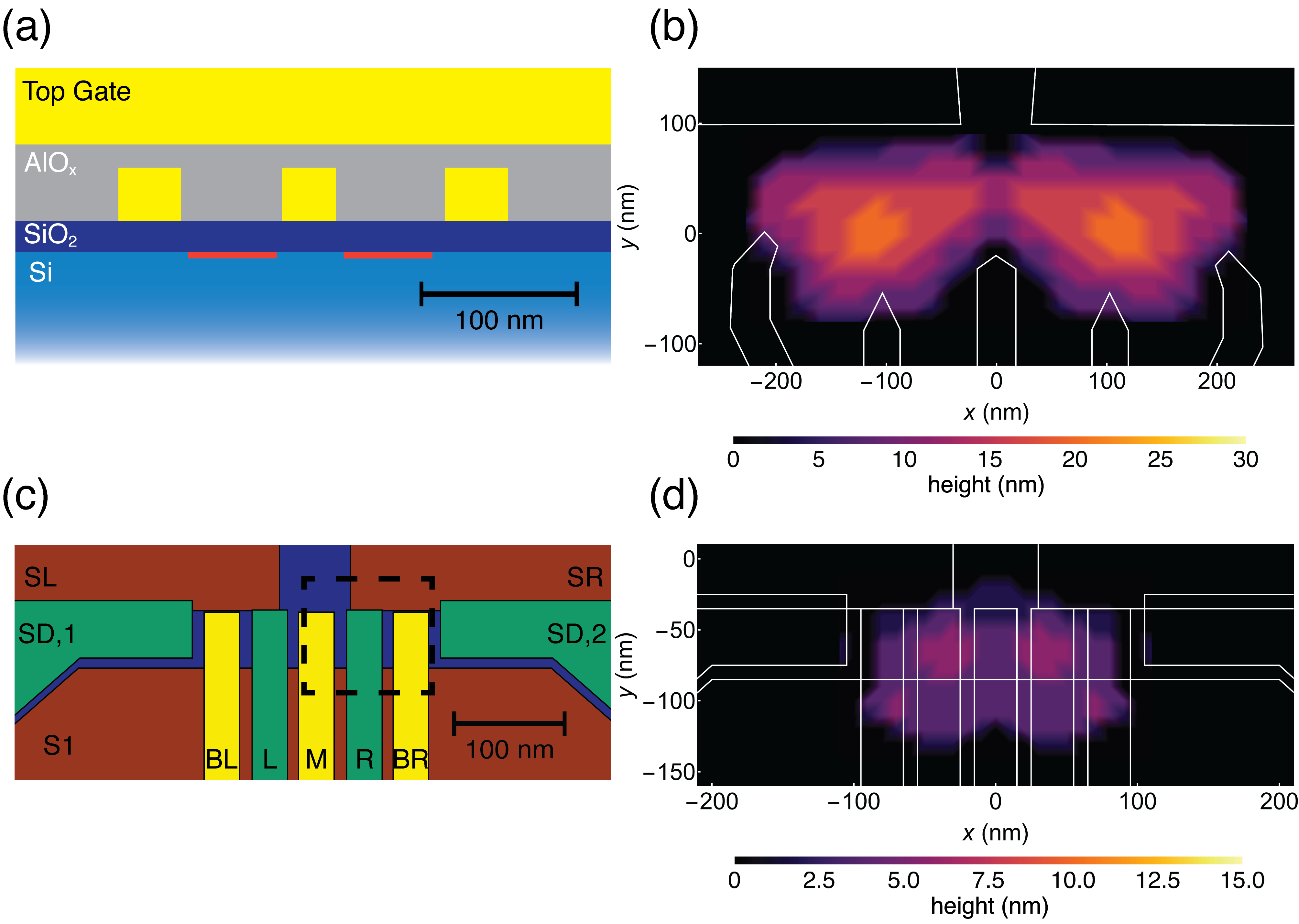}
	  \caption{Gate geometry modifications to reduce the effects of charged 
	  impurities on double quantum dot systems.
	  (a) Schematic illustrating reduced top gate to two-dimensional electron gas 
	  (2DEG) distance. Approximate dot locations indicated by red rectangles.
	  (b) Map of height above the 2DEG of the dangerous region to place a 
	  positively charged impurity.
	  The overall features are roughly similar to Fig.~\ref{fig:danger_map}, but the maximum dangerous height is reduced by half.  
	  (c) Schematic top view of an overlapping gate design.
	  The device heterostructure is the same as prior simulations.
	  Red gates are screening gates (SL, SR, S1), green gates are operated in 
	  accumulation mode to define the dots and reservoirs (SD,1; SD,2; L; R), and 
	  yellow gates are operated in depletion mode to create barriers (BL, M, BR). 
	  The dashed black rectangle indicates the lateral region used for the 
	  impurity studies. 
	  (d) Map of height above the 2DEG for dangerous positively charged impurity 
	  locations for an overlapping gate design device. This design exhibits both 
	  a sharply decreased spread of dangerous impurity sites as well as a maximum 
	  dangerous height one-sixth of that shown in Fig.~\ref{fig:danger_map}.
	  \label{fig:newideas}}
	\end{figure}

  We examined two modifications to the device design: (i) one in which the top 
  gate was moved closer to the 2DEG, reducing the separation between the global 
  top gate from  \SI{100}{\nano\metre} to \SI{50}{\nano\metre}, and (ii) the 
	other consisted of replacing the stadium style gates by an overlapping gate 
	design similar to that used in Ref.~\onlinecite{Zajac:2016fh}. 
	This style of design consists of three layers of gates: the first is a 
	screening layer that defines the dot channels, the second layer consists of 
	depletion gates, and the third layer consists of accumulation gates. 
	Separating the layers is a \SI{5}{\nano\metre} conformal layer of Al$_2$O$_3$.
	The devices we simulate are shown in Fig.\,\ref{fig:newideas}.
	Both changes would increase screening of charged impurities in the oxide and 
	are expected to increase the minimum impurity density to lift spin blockade. 
	  
	For the close-proximity top gate simulation, we considered charged impurities 
	in the same region as for the calculations summarized by Fig.\,3 of the main 
	text.
	For the analysis of the overlapping gate geometry, we considered a box with 
	dimensions, $\SI{110}{\nano\metre} \times \SI{170}{\nano\metre} \times 
	\SI{20}{\nano\metre}$ centered under the right accumulation gate starting at 
	the 2DEG interface and going up into the oxide, with a total of 1120 charge 
	locations, as indicated in Fig.\,\ref{fig:newideas}(c). 
	The overlapping gate design defines smaller dots and typically has a larger 
	orbital energy splitting than the stadium style designs.
	The energy cutoff for this design was chosen to be 
	\SI{7}{\milli\electronvolt}, approximately twice the orbital energy splitting 
	measured in Ref.~\onlinecite{Zajac:2016fh}. 
	Gate voltages that yield dots with single occupancy for the different
	designs are tabulated in Appendix~\ref{appendix:gate_voltage_tables}.
	  
	In the following, we consider an impurity location dangerous if an impurity 
	at that location leads to the spin blockade lifting effects discussed in the 
	main text.
	In the closer top gate design modification, the lateral features of the 
	dangerous region are largely unchanged, but the maximum height of that region 
	is reduced by a factor of two.
	The volume of dangerous impurity locations is \SI{7.1e5}{\nano\metre\cubed}, 
	as shown in Fig.\,\ref{fig:newideas}(b), which represents a slight 
	improvement on the original device, and a uniform impurity density of 
	\SI{1.4e15}{\per\centi\metre\cubed} would yield a high probability of finding 
	one impurity within this spin-blockade lifting volume.
	Considering only the charges at the interface, the area of dangerous impurity 
	locations is \SI{8.7e4}{\nano\metre\squared}.
	With a single charge within this area, the dangerous impurity density at the 
	interface is \SI{1.15e9}{\per\centi\metre\squared}.
	This increase in surface density represents a slight improvement on the 
	original device. 
	  
	The overlapping gate design greatly reduced the dangerous region.
	This can be attributed to the increased screening as well as the more tightly 
	confined dots inherent to this closely packed gate design, as this close 
	confinement reduces the wavefunction overlap between the lithographic dot and 
	the impurity level.
	The volume of spin-blockade lifting impurity locations, shown in 
	Fig.\,\ref{fig:newideas}(d), is \SI{6.3e4}{\nano\meter\cubed}, and the 
	impurity number density corresponding to one impurity within this volume is 
	\SI{1.6e16}{\per\centi\metre\cubed}.
	This increase in density represents a factor of eighteen improvement in 
	densities at which we would expect to start seeing spin blockade lifting 
	effects. 
	Considering only charges at the interface, the dangerous impurity area is 
	\SI{3.2e4}{\nano\metre\squared} and the corresponding dangerous surface 
	density is \SI{3.1e9}{\per\centi\metre\squared}.
	This limit on the interface impurity density, while improved from the devices 
	described in Fig.\,3 of the main text, is still lower than the expected 
	impurity density \cite{Sakaki:1970ii, Kim:2017hh} indicating that we would 
	still expect to see spin blockade lifting effects in these devices. 
	  
	  
	\bibliographystyle{apsrev4-1}
	\bibliography{UCLAPaperSources}

\begin{thebibliography}{37}%
\makeatletter
\providecommand \@ifxundefined [1]{%
 \@ifx{#1\undefined}
}%
\providecommand \@ifnum [1]{%
 \ifnum #1\expandafter \@firstoftwo
 \else \expandafter \@secondoftwo
 \fi
}%
\providecommand \@ifx [1]{%
 \ifx #1\expandafter \@firstoftwo
 \else \expandafter \@secondoftwo
 \fi
}%
\providecommand \natexlab [1]{#1}%
\providecommand \enquote  [1]{``#1''}%
\providecommand \bibnamefont  [1]{#1}%
\providecommand \bibfnamefont [1]{#1}%
\providecommand \citenamefont [1]{#1}%
\providecommand \href@noop [0]{\@secondoftwo}%
\providecommand \href [0]{\begingroup \@sanitize@url \@href}%
\providecommand \@href[1]{\@@startlink{#1}\@@href}%
\providecommand \@@href[1]{\endgroup#1\@@endlink}%
\providecommand \@sanitize@url [0]{\catcode `\\12\catcode `\$12\catcode
  `\&12\catcode `\#12\catcode `\^12\catcode `\_12\catcode `\%12\relax}%
\providecommand \@@startlink[1]{}%
\providecommand \@@endlink[0]{}%
\providecommand \url  [0]{\begingroup\@sanitize@url \@url }%
\providecommand \@url [1]{\endgroup\@href {#1}{\urlprefix }}%
\providecommand \urlprefix  [0]{URL }%
\providecommand \Eprint [0]{\href }%
\providecommand \doibase [0]{http://dx.doi.org/}%
\providecommand \selectlanguage [0]{\@gobble}%
\providecommand \bibinfo  [0]{\@secondoftwo}%
\providecommand \bibfield  [0]{\@secondoftwo}%
\providecommand \translation [1]{[#1]}%
\providecommand \BibitemOpen [0]{}%
\providecommand \bibitemStop [0]{}%
\providecommand \bibitemNoStop [0]{.\EOS\space}%
\providecommand \EOS [0]{\spacefactor3000\relax}%
\providecommand \BibitemShut  [1]{\csname bibitem#1\endcsname}%
\let\auto@bib@innerbib\@empty
\bibitem [{\citenamefont {Loss}\ and\ \citenamefont
  {DiVincenzo}(1998)}]{Loss:1998ia}%
  \BibitemOpen
  \bibfield  {author} {\bibinfo {author} {\bibfnamefont {D.}~\bibnamefont
  {Loss}}\ and\ \bibinfo {author} {\bibfnamefont {D.~P.}\ \bibnamefont
  {DiVincenzo}},\ }\href {\doibase 10.1103/PhysRevA.57.120} {\bibfield
  {journal} {\bibinfo  {journal} {Phys. Rev. A}\ }\textbf {\bibinfo {volume}
  {57}},\ \bibinfo {pages} {120} (\bibinfo {year} {1998})}\BibitemShut
  {NoStop}%
\bibitem [{\citenamefont {Zwanenburg}\ \emph {et~al.}(2013)\citenamefont
  {Zwanenburg}, \citenamefont {Dzurak}, \citenamefont {Morello}, \citenamefont
  {Simmons}, \citenamefont {Hollenberg}, \citenamefont {Klimeck}, \citenamefont
  {Rogge}, \citenamefont {Coppersmith},\ and\ \citenamefont
  {Eriksson}}]{Zwanenburg:2013gla}%
  \BibitemOpen
  \bibfield  {author} {\bibinfo {author} {\bibfnamefont {F.~A.}\ \bibnamefont
  {Zwanenburg}}, \bibinfo {author} {\bibfnamefont {A.~S.}\ \bibnamefont
  {Dzurak}}, \bibinfo {author} {\bibfnamefont {A.}~\bibnamefont {Morello}},
  \bibinfo {author} {\bibfnamefont {M.~Y.}\ \bibnamefont {Simmons}}, \bibinfo
  {author} {\bibfnamefont {L.~C.~L.}\ \bibnamefont {Hollenberg}}, \bibinfo
  {author} {\bibfnamefont {G.}~\bibnamefont {Klimeck}}, \bibinfo {author}
  {\bibfnamefont {S.}~\bibnamefont {Rogge}}, \bibinfo {author} {\bibfnamefont
  {S.~N.}\ \bibnamefont {Coppersmith}}, \ and\ \bibinfo {author} {\bibfnamefont
  {M.~A.}\ \bibnamefont {Eriksson}},\ }\href {\doibase
  10.1103/RevModPhys.85.961} {\bibfield  {journal} {\bibinfo  {journal} {Rev.
  Mod. Phys.}\ }\textbf {\bibinfo {volume} {85}},\ \bibinfo {pages} {961}
  (\bibinfo {year} {2013})}\BibitemShut {NoStop}%
\bibitem [{\citenamefont {Tyryshkin}\ \emph {et~al.}(2012)\citenamefont
  {Tyryshkin}, \citenamefont {Tojo}, \citenamefont {Morton}, \citenamefont
  {Riemann}, \citenamefont {Abrosimov}, \citenamefont {Becker}, \citenamefont
  {Pohl}, \citenamefont {Schenkel}, \citenamefont {Thewalt}, \citenamefont
  {Itoh},\ and\ \citenamefont {Lyon}}]{Tyryshkin:2012fi}%
  \BibitemOpen
  \bibfield  {author} {\bibinfo {author} {\bibfnamefont {A.~M.}\ \bibnamefont
  {Tyryshkin}}, \bibinfo {author} {\bibfnamefont {S.}~\bibnamefont {Tojo}},
  \bibinfo {author} {\bibfnamefont {J.~J.~L.}\ \bibnamefont {Morton}}, \bibinfo
  {author} {\bibfnamefont {H.}~\bibnamefont {Riemann}}, \bibinfo {author}
  {\bibfnamefont {N.~V.}\ \bibnamefont {Abrosimov}}, \bibinfo {author}
  {\bibfnamefont {P.}~\bibnamefont {Becker}}, \bibinfo {author} {\bibfnamefont
  {H.-J.}\ \bibnamefont {Pohl}}, \bibinfo {author} {\bibfnamefont
  {T.}~\bibnamefont {Schenkel}}, \bibinfo {author} {\bibfnamefont {M.~L.~W.}\
  \bibnamefont {Thewalt}}, \bibinfo {author} {\bibfnamefont {K.~M.}\
  \bibnamefont {Itoh}}, \ and\ \bibinfo {author} {\bibfnamefont {S.~A.}\
  \bibnamefont {Lyon}},\ }\href {\doibase 10.1038/nmat3182} {\bibfield
  {journal} {\bibinfo  {journal} {Nat. Mater.}\ }\textbf {\bibinfo {volume}
  {11}},\ \bibinfo {pages} {143} (\bibinfo {year} {2012})}\BibitemShut
  {NoStop}%
\bibitem [{\citenamefont {Veldhorst}\ \emph {et~al.}(2014)\citenamefont
  {Veldhorst}, \citenamefont {Hwang}, \citenamefont {Yang}, \citenamefont
  {Leenstra}, \citenamefont {de~Ronde}, \citenamefont {Dehollain},
  \citenamefont {Muhonen}, \citenamefont {Hudson}, \citenamefont {Itoh},
  \citenamefont {Morello},\ and\ \citenamefont {Dzurak}}]{Veldhorst:2014eq}%
  \BibitemOpen
  \bibfield  {author} {\bibinfo {author} {\bibfnamefont {M.}~\bibnamefont
  {Veldhorst}}, \bibinfo {author} {\bibfnamefont {J.~C.~C.}\ \bibnamefont
  {Hwang}}, \bibinfo {author} {\bibfnamefont {C.~H.}\ \bibnamefont {Yang}},
  \bibinfo {author} {\bibfnamefont {A.~W.}\ \bibnamefont {Leenstra}}, \bibinfo
  {author} {\bibfnamefont {B.}~\bibnamefont {de~Ronde}}, \bibinfo {author}
  {\bibfnamefont {J.~P.}\ \bibnamefont {Dehollain}}, \bibinfo {author}
  {\bibfnamefont {J.~T.}\ \bibnamefont {Muhonen}}, \bibinfo {author}
  {\bibfnamefont {F.~E.}\ \bibnamefont {Hudson}}, \bibinfo {author}
  {\bibfnamefont {K.~M.}\ \bibnamefont {Itoh}}, \bibinfo {author}
  {\bibfnamefont {A.}~\bibnamefont {Morello}}, \ and\ \bibinfo {author}
  {\bibfnamefont {A.~S.}\ \bibnamefont {Dzurak}},\ }\href {\doibase
  10.1038/nnano.2014.216} {\bibfield  {journal} {\bibinfo  {journal} {Nat.
  Nanotechnol.}\ }\textbf {\bibinfo {volume} {9}},\ \bibinfo {pages} {981}
  (\bibinfo {year} {2014})}\BibitemShut {NoStop}%
\bibitem [{\citenamefont {Veldhorst}\ \emph {et~al.}(2015)\citenamefont
  {Veldhorst}, \citenamefont {Yang}, \citenamefont {Hwang}, \citenamefont
  {Huang}, \citenamefont {Dehollain}, \citenamefont {Muhonen}, \citenamefont
  {Simmons}, \citenamefont {Laucht}, \citenamefont {Hudson}, \citenamefont
  {Itoh}, \citenamefont {Morello},\ and\ \citenamefont
  {Dzurak}}]{Veldhorst:2015je}%
  \BibitemOpen
  \bibfield  {author} {\bibinfo {author} {\bibfnamefont {M.}~\bibnamefont
  {Veldhorst}}, \bibinfo {author} {\bibfnamefont {C.~H.}\ \bibnamefont {Yang}},
  \bibinfo {author} {\bibfnamefont {J.~C.~C.}\ \bibnamefont {Hwang}}, \bibinfo
  {author} {\bibfnamefont {W.}~\bibnamefont {Huang}}, \bibinfo {author}
  {\bibfnamefont {J.~P.}\ \bibnamefont {Dehollain}}, \bibinfo {author}
  {\bibfnamefont {J.~T.}\ \bibnamefont {Muhonen}}, \bibinfo {author}
  {\bibfnamefont {S.}~\bibnamefont {Simmons}}, \bibinfo {author} {\bibfnamefont
  {A.}~\bibnamefont {Laucht}}, \bibinfo {author} {\bibfnamefont {F.~E.}\
  \bibnamefont {Hudson}}, \bibinfo {author} {\bibfnamefont {K.~M.}\
  \bibnamefont {Itoh}}, \bibinfo {author} {\bibfnamefont {A.}~\bibnamefont
  {Morello}}, \ and\ \bibinfo {author} {\bibfnamefont {A.~S.}\ \bibnamefont
  {Dzurak}},\ }\href {\doibase 10.1038/nature15263} {\bibfield  {journal}
  {\bibinfo  {journal} {Nature}\ }\textbf {\bibinfo {volume} {526}},\ \bibinfo
  {pages} {410} (\bibinfo {year} {2015})}\BibitemShut {NoStop}%
\bibitem [{\citenamefont {Jock}\ \emph {et~al.}(2018)\citenamefont {Jock},
  \citenamefont {Jacobson}, \citenamefont {Harvey-Collard}, \citenamefont
  {Mounce}, \citenamefont {Srinivasa}, \citenamefont {Ward}, \citenamefont
  {Anderson}, \citenamefont {Manginell}, \citenamefont {Wendt}, \citenamefont
  {Rudolph}, \citenamefont {Pluym}, \citenamefont {Gamble}, \citenamefont
  {Baczewski}, \citenamefont {Witzel},\ and\ \citenamefont
  {Carroll}}]{Jock:2018cu}%
  \BibitemOpen
  \bibfield  {author} {\bibinfo {author} {\bibfnamefont {R.~M.}\ \bibnamefont
  {Jock}}, \bibinfo {author} {\bibfnamefont {N.~T.}\ \bibnamefont {Jacobson}},
  \bibinfo {author} {\bibfnamefont {P.}~\bibnamefont {Harvey-Collard}},
  \bibinfo {author} {\bibfnamefont {A.~M.}\ \bibnamefont {Mounce}}, \bibinfo
  {author} {\bibfnamefont {V.}~\bibnamefont {Srinivasa}}, \bibinfo {author}
  {\bibfnamefont {D.~R.}\ \bibnamefont {Ward}}, \bibinfo {author}
  {\bibfnamefont {J.}~\bibnamefont {Anderson}}, \bibinfo {author}
  {\bibfnamefont {R.}~\bibnamefont {Manginell}}, \bibinfo {author}
  {\bibfnamefont {J.~R.}\ \bibnamefont {Wendt}}, \bibinfo {author}
  {\bibfnamefont {M.}~\bibnamefont {Rudolph}}, \bibinfo {author} {\bibfnamefont
  {T.}~\bibnamefont {Pluym}}, \bibinfo {author} {\bibfnamefont {J.~K.}\
  \bibnamefont {Gamble}}, \bibinfo {author} {\bibfnamefont {A.~D.}\
  \bibnamefont {Baczewski}}, \bibinfo {author} {\bibfnamefont {W.~M.}\
  \bibnamefont {Witzel}}, \ and\ \bibinfo {author} {\bibfnamefont {M.~S.}\
  \bibnamefont {Carroll}},\ }\href {\doibase 10.1038/s41467-018-04200-0}
  {\bibfield  {journal} {\bibinfo  {journal} {Nat. Comm.}\ }\textbf {\bibinfo
  {volume} {9}},\ \bibinfo {pages} {1768} (\bibinfo {year} {2018})}\BibitemShut
  {NoStop}%
\bibitem [{\citenamefont {Levy}(2002)}]{Levy:2002ex}%
  \BibitemOpen
  \bibfield  {author} {\bibinfo {author} {\bibfnamefont {J.}~\bibnamefont
  {Levy}},\ }\href {\doibase 10.1103/PhysRevLett.89.147902} {\bibfield
  {journal} {\bibinfo  {journal} {Phys. Rev. Lett.}\ }\textbf {\bibinfo
  {volume} {89}},\ \bibinfo {pages} {147902} (\bibinfo {year}
  {2002})}\BibitemShut {NoStop}%
\bibitem [{\citenamefont {Petta}\ \emph {et~al.}(2005)\citenamefont {Petta},
  \citenamefont {Johnson}, \citenamefont {Taylor}, \citenamefont {Laird},
  \citenamefont {Yacoby}, \citenamefont {Lukin}, \citenamefont {Marcus},
  \citenamefont {Hanson},\ and\ \citenamefont {Gossard}}]{Petta:2005kn}%
  \BibitemOpen
  \bibfield  {author} {\bibinfo {author} {\bibfnamefont {J.~R.}\ \bibnamefont
  {Petta}}, \bibinfo {author} {\bibfnamefont {A.~C.}\ \bibnamefont {Johnson}},
  \bibinfo {author} {\bibfnamefont {J.~M.}\ \bibnamefont {Taylor}}, \bibinfo
  {author} {\bibfnamefont {E.~A.}\ \bibnamefont {Laird}}, \bibinfo {author}
  {\bibfnamefont {A.}~\bibnamefont {Yacoby}}, \bibinfo {author} {\bibfnamefont
  {M.~D.}\ \bibnamefont {Lukin}}, \bibinfo {author} {\bibfnamefont {C.~M.}\
  \bibnamefont {Marcus}}, \bibinfo {author} {\bibfnamefont {M.~P.}\
  \bibnamefont {Hanson}}, \ and\ \bibinfo {author} {\bibfnamefont {A.~C.}\
  \bibnamefont {Gossard}},\ }\href {\doibase 10.1126/science.1116955}
  {\bibfield  {journal} {\bibinfo  {journal} {Science}\ }\textbf {\bibinfo
  {volume} {309}},\ \bibinfo {pages} {2180} (\bibinfo {year}
  {2005})}\BibitemShut {NoStop}%
\bibitem [{\citenamefont {Ciorga}\ \emph {et~al.}(2000)\citenamefont {Ciorga},
  \citenamefont {Sachrajda}, \citenamefont {Hawrylak}, \citenamefont {Gould},
  \citenamefont {Zawadzki}, \citenamefont {Jullian}, \citenamefont {Feng},\
  and\ \citenamefont {Wasilewski}}]{Ciorga:2000br}%
  \BibitemOpen
  \bibfield  {author} {\bibinfo {author} {\bibfnamefont {M.}~\bibnamefont
  {Ciorga}}, \bibinfo {author} {\bibfnamefont {A.~S.}\ \bibnamefont
  {Sachrajda}}, \bibinfo {author} {\bibfnamefont {P.}~\bibnamefont {Hawrylak}},
  \bibinfo {author} {\bibfnamefont {C.}~\bibnamefont {Gould}}, \bibinfo
  {author} {\bibfnamefont {P.}~\bibnamefont {Zawadzki}}, \bibinfo {author}
  {\bibfnamefont {S.}~\bibnamefont {Jullian}}, \bibinfo {author} {\bibfnamefont
  {Y.}~\bibnamefont {Feng}}, \ and\ \bibinfo {author} {\bibfnamefont
  {Z.}~\bibnamefont {Wasilewski}},\ }\href {\doibase
  10.1103/PhysRevB.61.R16315} {\bibfield  {journal} {\bibinfo  {journal} {Phys.
  Rev. B}\ }\textbf {\bibinfo {volume} {61}},\ \bibinfo {pages} {R16315}
  (\bibinfo {year} {2000})}\BibitemShut {NoStop}%
\bibitem [{\citenamefont {Ono}\ \emph {et~al.}(2002)\citenamefont {Ono},
  \citenamefont {Austing}, \citenamefont {Tokura},\ and\ \citenamefont
  {Tarucha}}]{Ono:2002gz}%
  \BibitemOpen
  \bibfield  {author} {\bibinfo {author} {\bibfnamefont {K.}~\bibnamefont
  {Ono}}, \bibinfo {author} {\bibfnamefont {D.~G.}\ \bibnamefont {Austing}},
  \bibinfo {author} {\bibfnamefont {Y.}~\bibnamefont {Tokura}}, \ and\ \bibinfo
  {author} {\bibfnamefont {S.}~\bibnamefont {Tarucha}},\ }\href {\doibase
  10.1126/science.1070958} {\bibfield  {journal} {\bibinfo  {journal}
  {Science}\ }\textbf {\bibinfo {volume} {297}},\ \bibinfo {pages} {1313}
  (\bibinfo {year} {2002})}\BibitemShut {NoStop}%
\bibitem [{\citenamefont {Hao}\ \emph {et~al.}(2014)\citenamefont {Hao},
  \citenamefont {Ruskov}, \citenamefont {Xiao}, \citenamefont {Tahan},\ and\
  \citenamefont {Jiang}}]{Hao:2014ea}%
  \BibitemOpen
  \bibfield  {author} {\bibinfo {author} {\bibfnamefont {X.}~\bibnamefont
  {Hao}}, \bibinfo {author} {\bibfnamefont {R.}~\bibnamefont {Ruskov}},
  \bibinfo {author} {\bibfnamefont {M.}~\bibnamefont {Xiao}}, \bibinfo {author}
  {\bibfnamefont {C.}~\bibnamefont {Tahan}}, \ and\ \bibinfo {author}
  {\bibfnamefont {H.}~\bibnamefont {Jiang}},\ }\href {\doibase
  10.1038/ncomms4860} {\bibfield  {journal} {\bibinfo  {journal} {Nat. Comm.}\
  }\textbf {\bibinfo {volume} {5}},\ \bibinfo {pages} {344} (\bibinfo {year}
  {2014})}\BibitemShut {NoStop}%
\bibitem [{\citenamefont {Maune}\ \emph {et~al.}(2012)\citenamefont {Maune},
  \citenamefont {Borselli}, \citenamefont {Huang}, \citenamefont {Ladd},
  \citenamefont {Deelman}, \citenamefont {Holabird}, \citenamefont {Kiselev},
  \citenamefont {Alvarado-Rodriguez}, \citenamefont {Ross}, \citenamefont
  {Schmitz}, \citenamefont {Sokolich}, \citenamefont {Watson}, \citenamefont
  {Gyure},\ and\ \citenamefont {Hunter}}]{Maune:2012iu}%
  \BibitemOpen
  \bibfield  {author} {\bibinfo {author} {\bibfnamefont {B.~M.}\ \bibnamefont
  {Maune}}, \bibinfo {author} {\bibfnamefont {M.~G.}\ \bibnamefont {Borselli}},
  \bibinfo {author} {\bibfnamefont {B.}~\bibnamefont {Huang}}, \bibinfo
  {author} {\bibfnamefont {T.~D.}\ \bibnamefont {Ladd}}, \bibinfo {author}
  {\bibfnamefont {P.~W.}\ \bibnamefont {Deelman}}, \bibinfo {author}
  {\bibfnamefont {K.~S.}\ \bibnamefont {Holabird}}, \bibinfo {author}
  {\bibfnamefont {A.~A.}\ \bibnamefont {Kiselev}}, \bibinfo {author}
  {\bibfnamefont {I.}~\bibnamefont {Alvarado-Rodriguez}}, \bibinfo {author}
  {\bibfnamefont {R.~S.}\ \bibnamefont {Ross}}, \bibinfo {author}
  {\bibfnamefont {A.~E.}\ \bibnamefont {Schmitz}}, \bibinfo {author}
  {\bibfnamefont {M.}~\bibnamefont {Sokolich}}, \bibinfo {author}
  {\bibfnamefont {C.~A.}\ \bibnamefont {Watson}}, \bibinfo {author}
  {\bibfnamefont {M.~F.}\ \bibnamefont {Gyure}}, \ and\ \bibinfo {author}
  {\bibfnamefont {A.~T.}\ \bibnamefont {Hunter}},\ }\href {\doibase
  10.1038/nature10707} {\bibfield  {journal} {\bibinfo  {journal} {Nature}\
  }\textbf {\bibinfo {volume} {481}},\ \bibinfo {pages} {344} (\bibinfo {year}
  {2012})}\BibitemShut {NoStop}%
\bibitem [{\citenamefont {Wu}\ \emph {et~al.}(2014)\citenamefont {Wu},
  \citenamefont {Ward}, \citenamefont {Prance}, \citenamefont {Kim},
  \citenamefont {Gamble}, \citenamefont {Mohr}, \citenamefont {Shi},
  \citenamefont {Savage}, \citenamefont {Lagally}, \citenamefont {Friesen},
  \citenamefont {Coppersmith},\ and\ \citenamefont {Eriksson}}]{Wu:2014fz}%
  \BibitemOpen
  \bibfield  {author} {\bibinfo {author} {\bibfnamefont {X.}~\bibnamefont
  {Wu}}, \bibinfo {author} {\bibfnamefont {D.~R.}\ \bibnamefont {Ward}},
  \bibinfo {author} {\bibfnamefont {J.~R.}\ \bibnamefont {Prance}}, \bibinfo
  {author} {\bibfnamefont {D.}~\bibnamefont {Kim}}, \bibinfo {author}
  {\bibfnamefont {J.~K.}\ \bibnamefont {Gamble}}, \bibinfo {author}
  {\bibfnamefont {R.~T.}\ \bibnamefont {Mohr}}, \bibinfo {author}
  {\bibfnamefont {Z.}~\bibnamefont {Shi}}, \bibinfo {author} {\bibfnamefont
  {D.~E.}\ \bibnamefont {Savage}}, \bibinfo {author} {\bibfnamefont {M.~G.}\
  \bibnamefont {Lagally}}, \bibinfo {author} {\bibfnamefont {M.}~\bibnamefont
  {Friesen}}, \bibinfo {author} {\bibfnamefont {S.~N.}\ \bibnamefont
  {Coppersmith}}, \ and\ \bibinfo {author} {\bibfnamefont {M.~A.}\ \bibnamefont
  {Eriksson}},\ }\href {\doibase 10.1073/pnas.1412230111} {\bibfield  {journal}
  {\bibinfo  {journal} {Proc. Natl. Acad. Sci. {U.S.A.}}\ }\textbf {\bibinfo
  {volume} {111}},\ \bibinfo {pages} {11938} (\bibinfo {year}
  {2014})}\BibitemShut {NoStop}%
\bibitem [{\citenamefont {Harvey-Collard}\ \emph {et~al.}(2017)\citenamefont
  {Harvey-Collard}, \citenamefont {Jacobson}, \citenamefont {Rudolph},
  \citenamefont {Dominguez}, \citenamefont {Ten~Eyck}, \citenamefont {Wendt},
  \citenamefont {Pluym}, \citenamefont {Gamble}, \citenamefont {Lilly},
  \citenamefont {Pioro-Ladri{\`e}re},\ and\ \citenamefont
  {Carroll}}]{HarveyCollard:2017ic}%
  \BibitemOpen
  \bibfield  {author} {\bibinfo {author} {\bibfnamefont {P.}~\bibnamefont
  {Harvey-Collard}}, \bibinfo {author} {\bibfnamefont {N.~T.}\ \bibnamefont
  {Jacobson}}, \bibinfo {author} {\bibfnamefont {M.}~\bibnamefont {Rudolph}},
  \bibinfo {author} {\bibfnamefont {J.}~\bibnamefont {Dominguez}}, \bibinfo
  {author} {\bibfnamefont {G.~A.}\ \bibnamefont {Ten~Eyck}}, \bibinfo {author}
  {\bibfnamefont {J.~R.}\ \bibnamefont {Wendt}}, \bibinfo {author}
  {\bibfnamefont {T.}~\bibnamefont {Pluym}}, \bibinfo {author} {\bibfnamefont
  {J.~K.}\ \bibnamefont {Gamble}}, \bibinfo {author} {\bibfnamefont {M.~P.}\
  \bibnamefont {Lilly}}, \bibinfo {author} {\bibfnamefont {M.}~\bibnamefont
  {Pioro-Ladri{\`e}re}}, \ and\ \bibinfo {author} {\bibfnamefont {M.~S.}\
  \bibnamefont {Carroll}},\ }\href {\doibase 10.1038/s41467-017-01113-2}
  {\bibfield  {journal} {\bibinfo  {journal} {Nat. Comm.}\ }\textbf {\bibinfo
  {volume} {8}},\ \bibinfo {pages} {1029} (\bibinfo {year} {2017})}\BibitemShut
  {NoStop}%
\bibitem [{\citenamefont {Veldhorst}\ \emph {et~al.}(2017)\citenamefont
  {Veldhorst}, \citenamefont {Eenink}, \citenamefont {Yang},\ and\
  \citenamefont {Dzurak}}]{Veldhorst:2017ht}%
  \BibitemOpen
  \bibfield  {author} {\bibinfo {author} {\bibfnamefont {M.}~\bibnamefont
  {Veldhorst}}, \bibinfo {author} {\bibfnamefont {H.~G.~J.}\ \bibnamefont
  {Eenink}}, \bibinfo {author} {\bibfnamefont {C.~H.}\ \bibnamefont {Yang}}, \
  and\ \bibinfo {author} {\bibfnamefont {A.~S.}\ \bibnamefont {Dzurak}},\
  }\href {\doibase 10.1038/s41467-017-01905-6} {\bibfield  {journal} {\bibinfo
  {journal} {Nature Comm.}\ }\textbf {\bibinfo {volume} {8}},\ \bibinfo {pages}
  {1766} (\bibinfo {year} {2017})}\BibitemShut {NoStop}%
\bibitem [{\citenamefont {Xiao}\ \emph {et~al.}(2010)\citenamefont {Xiao},
  \citenamefont {House},\ and\ \citenamefont {Jiang}}]{Xiao:2010cx}%
  \BibitemOpen
  \bibfield  {author} {\bibinfo {author} {\bibfnamefont {M.}~\bibnamefont
  {Xiao}}, \bibinfo {author} {\bibfnamefont {M.~G.}\ \bibnamefont {House}}, \
  and\ \bibinfo {author} {\bibfnamefont {H.~W.}\ \bibnamefont {Jiang}},\ }\href
  {\doibase 10.1103/PhysRevLett.104.096801} {\bibfield  {journal} {\bibinfo
  {journal} {Phys. Rev. Lett.}\ }\textbf {\bibinfo {volume} {104}},\ \bibinfo
  {pages} {096801} (\bibinfo {year} {2010})}\BibitemShut {NoStop}%
\bibitem [{\citenamefont {Maurand}\ \emph {et~al.}(2016)\citenamefont
  {Maurand}, \citenamefont {Jehl}, \citenamefont {Kotekar-Patil}, \citenamefont
  {Corna}, \citenamefont {Bohuslavskyi}, \citenamefont {Lavi{\'e}ville},
  \citenamefont {Hutin}, \citenamefont {Barraud}, \citenamefont {Vinet},
  \citenamefont {Sanquer},\ and\ \citenamefont
  {De~Franceschi}}]{Maurand:2016cj}%
  \BibitemOpen
  \bibfield  {author} {\bibinfo {author} {\bibfnamefont {R.}~\bibnamefont
  {Maurand}}, \bibinfo {author} {\bibfnamefont {X.}~\bibnamefont {Jehl}},
  \bibinfo {author} {\bibfnamefont {D.}~\bibnamefont {Kotekar-Patil}}, \bibinfo
  {author} {\bibfnamefont {A.}~\bibnamefont {Corna}}, \bibinfo {author}
  {\bibfnamefont {H.}~\bibnamefont {Bohuslavskyi}}, \bibinfo {author}
  {\bibfnamefont {R.}~\bibnamefont {Lavi{\'e}ville}}, \bibinfo {author}
  {\bibfnamefont {L.}~\bibnamefont {Hutin}}, \bibinfo {author} {\bibfnamefont
  {S.}~\bibnamefont {Barraud}}, \bibinfo {author} {\bibfnamefont
  {M.}~\bibnamefont {Vinet}}, \bibinfo {author} {\bibfnamefont
  {M.}~\bibnamefont {Sanquer}}, \ and\ \bibinfo {author} {\bibfnamefont
  {S.}~\bibnamefont {De~Franceschi}},\ }\href {\doibase 10.1038/ncomms13575}
  {\bibfield  {journal} {\bibinfo  {journal} {Nat. Comm.}\ }\textbf {\bibinfo
  {volume} {7}},\ \bibinfo {pages} {13575} (\bibinfo {year}
  {2016})}\BibitemShut {NoStop}%
\bibitem [{\citenamefont {Zajac}\ \emph {et~al.}(2016)\citenamefont {Zajac},
  \citenamefont {Hazard}, \citenamefont {Mi}, \citenamefont {Nielsen},\ and\
  \citenamefont {Petta}}]{Zajac:2016fh}%
  \BibitemOpen
  \bibfield  {author} {\bibinfo {author} {\bibfnamefont {D.~M.}\ \bibnamefont
  {Zajac}}, \bibinfo {author} {\bibfnamefont {T.~M.}\ \bibnamefont {Hazard}},
  \bibinfo {author} {\bibfnamefont {X.}~\bibnamefont {Mi}}, \bibinfo {author}
  {\bibfnamefont {E.}~\bibnamefont {Nielsen}}, \ and\ \bibinfo {author}
  {\bibfnamefont {J.~R.}\ \bibnamefont {Petta}},\ }\href {\doibase
  10.1103/PhysRevApplied.6.054013} {\bibfield  {journal} {\bibinfo  {journal}
  {Phys. Rev. Applied}\ }\textbf {\bibinfo {volume} {6}},\ \bibinfo {pages}
  {054013} (\bibinfo {year} {2016})}\BibitemShut {NoStop}%
\bibitem [{\citenamefont {Shaji}\ \emph {et~al.}(2008)\citenamefont {Shaji},
  \citenamefont {Simmons}, \citenamefont {Thalakulam}, \citenamefont {Klein},
  \citenamefont {Qin}, \citenamefont {Luo}, \citenamefont {Savage},
  \citenamefont {Lagally}, \citenamefont {Rimberg}, \citenamefont {Joynt},
  \citenamefont {Friesen}, \citenamefont {Blick}, \citenamefont {Coppersmith},\
  and\ \citenamefont {Eriksson}}]{Shaji:2008hq}%
  \BibitemOpen
  \bibfield  {author} {\bibinfo {author} {\bibfnamefont {N.}~\bibnamefont
  {Shaji}}, \bibinfo {author} {\bibfnamefont {C.~B.}\ \bibnamefont {Simmons}},
  \bibinfo {author} {\bibfnamefont {M.}~\bibnamefont {Thalakulam}}, \bibinfo
  {author} {\bibfnamefont {L.~J.}\ \bibnamefont {Klein}}, \bibinfo {author}
  {\bibfnamefont {H.}~\bibnamefont {Qin}}, \bibinfo {author} {\bibfnamefont
  {H.}~\bibnamefont {Luo}}, \bibinfo {author} {\bibfnamefont {D.~E.}\
  \bibnamefont {Savage}}, \bibinfo {author} {\bibfnamefont {M.~G.}\
  \bibnamefont {Lagally}}, \bibinfo {author} {\bibfnamefont {A.~J.}\
  \bibnamefont {Rimberg}}, \bibinfo {author} {\bibfnamefont {R.}~\bibnamefont
  {Joynt}}, \bibinfo {author} {\bibfnamefont {M.}~\bibnamefont {Friesen}},
  \bibinfo {author} {\bibfnamefont {R.~H.}\ \bibnamefont {Blick}}, \bibinfo
  {author} {\bibfnamefont {S.~N.}\ \bibnamefont {Coppersmith}}, \ and\ \bibinfo
  {author} {\bibfnamefont {M.~A.}\ \bibnamefont {Eriksson}},\ }\href {\doibase
  10.1038/nphys988} {\bibfield  {journal} {\bibinfo  {journal} {Nat. Phys.}\
  }\textbf {\bibinfo {volume} {4}},\ \bibinfo {pages} {540} (\bibinfo {year}
  {2008})}\BibitemShut {NoStop}%
\bibitem [{\citenamefont {Simmons}\ \emph {et~al.}(2010)\citenamefont
  {Simmons}, \citenamefont {Koh}, \citenamefont {Shaji}, \citenamefont
  {Thalakulam}, \citenamefont {Klein}, \citenamefont {Qin}, \citenamefont
  {Luo}, \citenamefont {Savage}, \citenamefont {Lagally}, \citenamefont
  {Rimberg}, \citenamefont {Joynt}, \citenamefont {Blick}, \citenamefont
  {Friesen}, \citenamefont {Coppersmith},\ and\ \citenamefont
  {Eriksson}}]{Simmons:2010ev}%
  \BibitemOpen
  \bibfield  {author} {\bibinfo {author} {\bibfnamefont {C.~B.}\ \bibnamefont
  {Simmons}}, \bibinfo {author} {\bibfnamefont {T.~S.}\ \bibnamefont {Koh}},
  \bibinfo {author} {\bibfnamefont {N.}~\bibnamefont {Shaji}}, \bibinfo
  {author} {\bibfnamefont {M.}~\bibnamefont {Thalakulam}}, \bibinfo {author}
  {\bibfnamefont {L.~J.}\ \bibnamefont {Klein}}, \bibinfo {author}
  {\bibfnamefont {H.}~\bibnamefont {Qin}}, \bibinfo {author} {\bibfnamefont
  {H.}~\bibnamefont {Luo}}, \bibinfo {author} {\bibfnamefont {D.~E.}\
  \bibnamefont {Savage}}, \bibinfo {author} {\bibfnamefont {M.~G.}\
  \bibnamefont {Lagally}}, \bibinfo {author} {\bibfnamefont {A.~J.}\
  \bibnamefont {Rimberg}}, \bibinfo {author} {\bibfnamefont {R.}~\bibnamefont
  {Joynt}}, \bibinfo {author} {\bibfnamefont {R.}~\bibnamefont {Blick}},
  \bibinfo {author} {\bibfnamefont {M.}~\bibnamefont {Friesen}}, \bibinfo
  {author} {\bibfnamefont {S.~N.}\ \bibnamefont {Coppersmith}}, \ and\ \bibinfo
  {author} {\bibfnamefont {M.~A.}\ \bibnamefont {Eriksson}},\ }\href {\doibase
  10.1103/PhysRevB.82.245312} {\bibfield  {journal} {\bibinfo  {journal} {Phys.
  Rev. B}\ }\textbf {\bibinfo {volume} {82}},\ \bibinfo {pages} {245312}
  (\bibinfo {year} {2010})}\BibitemShut {NoStop}%
\bibitem [{\citenamefont {Koh}\ \emph {et~al.}(2011)\citenamefont {Koh},
  \citenamefont {Simmons}, \citenamefont {Eriksson}, \citenamefont
  {Coppersmith},\ and\ \citenamefont {Friesen}}]{Koh:2011cx}%
  \BibitemOpen
  \bibfield  {author} {\bibinfo {author} {\bibfnamefont {T.~S.}\ \bibnamefont
  {Koh}}, \bibinfo {author} {\bibfnamefont {C.~B.}\ \bibnamefont {Simmons}},
  \bibinfo {author} {\bibfnamefont {M.~A.}\ \bibnamefont {Eriksson}}, \bibinfo
  {author} {\bibfnamefont {S.~N.}\ \bibnamefont {Coppersmith}}, \ and\ \bibinfo
  {author} {\bibfnamefont {M.}~\bibnamefont {Friesen}},\ }\href {\doibase
  10.1103/PhysRevLett.106.186801} {\bibfield  {journal} {\bibinfo  {journal}
  {Phys. Rev. Lett.}\ }\textbf {\bibinfo {volume} {106}},\ \bibinfo {pages}
  {186801} (\bibinfo {year} {2011})}\BibitemShut {NoStop}%
\bibitem [{\citenamefont {Shi}\ \emph {et~al.}(2012)\citenamefont {Shi},
  \citenamefont {Simmons}, \citenamefont {Prance}, \citenamefont {Gamble},
  \citenamefont {Koh}, \citenamefont {Shim}, \citenamefont {Hu}, \citenamefont
  {Savage}, \citenamefont {Lagally}, \citenamefont {Eriksson}, \citenamefont
  {Friesen},\ and\ \citenamefont {Coppersmith}}]{Shi:2012kla}%
  \BibitemOpen
  \bibfield  {author} {\bibinfo {author} {\bibfnamefont {Z.}~\bibnamefont
  {Shi}}, \bibinfo {author} {\bibfnamefont {C.~B.}\ \bibnamefont {Simmons}},
  \bibinfo {author} {\bibfnamefont {J.~R.}\ \bibnamefont {Prance}}, \bibinfo
  {author} {\bibfnamefont {J.~K.}\ \bibnamefont {Gamble}}, \bibinfo {author}
  {\bibfnamefont {T.~S.}\ \bibnamefont {Koh}}, \bibinfo {author} {\bibfnamefont
  {Y.-P.}\ \bibnamefont {Shim}}, \bibinfo {author} {\bibfnamefont
  {X.}~\bibnamefont {Hu}}, \bibinfo {author} {\bibfnamefont {D.~E.}\
  \bibnamefont {Savage}}, \bibinfo {author} {\bibfnamefont {M.~G.}\
  \bibnamefont {Lagally}}, \bibinfo {author} {\bibfnamefont {M.~A.}\
  \bibnamefont {Eriksson}}, \bibinfo {author} {\bibfnamefont {M.}~\bibnamefont
  {Friesen}}, \ and\ \bibinfo {author} {\bibfnamefont {S.~N.}\ \bibnamefont
  {Coppersmith}},\ }\href {\doibase 10.1103/PhysRevLett.108.140503} {\bibfield
  {journal} {\bibinfo  {journal} {Phys. Rev. Lett.}\ }\textbf {\bibinfo
  {volume} {108}},\ \bibinfo {pages} {140503} (\bibinfo {year}
  {2012})}\BibitemShut {NoStop}%
\bibitem [{\citenamefont {Cao}\ \emph {et~al.}(2016)\citenamefont {Cao},
  \citenamefont {Li}, \citenamefont {Yu}, \citenamefont {Wang}, \citenamefont
  {Chen}, \citenamefont {Song}, \citenamefont {Xiao}, \citenamefont {Guo},
  \citenamefont {Jiang}, \citenamefont {Hu}, ,\ and\ \citenamefont
  {Guo}}]{cao2016tunable}%
  \BibitemOpen
  \bibfield  {author} {\bibinfo {author} {\bibfnamefont {G.}~\bibnamefont
  {Cao}}, \bibinfo {author} {\bibfnamefont {H.-O.}\ \bibnamefont {Li}},
  \bibinfo {author} {\bibfnamefont {G.-D.}\ \bibnamefont {Yu}}, \bibinfo
  {author} {\bibfnamefont {B.-C.}\ \bibnamefont {Wang}}, \bibinfo {author}
  {\bibfnamefont {B.-B.}\ \bibnamefont {Chen}}, \bibinfo {author}
  {\bibfnamefont {X.-X.}\ \bibnamefont {Song}}, \bibinfo {author}
  {\bibfnamefont {M.}~\bibnamefont {Xiao}}, \bibinfo {author} {\bibfnamefont
  {G.-C.}\ \bibnamefont {Guo}}, \bibinfo {author} {\bibfnamefont {H.-W.}\
  \bibnamefont {Jiang}}, \bibinfo {author} {\bibfnamefont {X.}~\bibnamefont
  {Hu}}, , \ and\ \bibinfo {author} {\bibfnamefont {G.-P.}\ \bibnamefont
  {Guo}},\ }\href@noop {} {\bibfield  {journal} {\bibinfo  {journal} {Physical
  Review Letters}\ }\textbf {\bibinfo {volume} {116}},\ \bibinfo {pages}
  {086801} (\bibinfo {year} {2016})}\BibitemShut {NoStop}%
\bibitem [{\citenamefont {Stopa}(1996)}]{Stopa:1996kr}%
  \BibitemOpen
  \bibfield  {author} {\bibinfo {author} {\bibfnamefont {M.}~\bibnamefont
  {Stopa}},\ }\href {\doibase 10.1103/PhysRevB.54.13767} {\bibfield  {journal}
  {\bibinfo  {journal} {Phys. Rev. B}\ }\textbf {\bibinfo {volume} {54}},\
  \bibinfo {pages} {13767} (\bibinfo {year} {1996})}\BibitemShut {NoStop}%
\bibitem [{com()}]{comsol}%
  \BibitemOpen
  \href {www.comsol.com} {}\bibinfo {note} {{COMSOL}
  Multiphysics\textsuperscript{\textregistered} v. 5.3a. www.comsol.com. COMSOL
  AB, Stockholm, Sweden}\BibitemShut {NoStop}%
\bibitem [{\citenamefont {Zhao}\ \emph {et~al.}(2019)\citenamefont {Zhao},
  \citenamefont {Tanttu}, \citenamefont {Tan}, \citenamefont {Hensen},
  \citenamefont {Chan}, \citenamefont {Hwang}, \citenamefont {Leon},
  \citenamefont {Yang}, \citenamefont {Gilbert}, \citenamefont {Hudson},
  \citenamefont {Itoh}, \citenamefont {Kiselev}, \citenamefont {Ladd},
  \citenamefont {Morello}, \citenamefont {Laucht},\ and\ \citenamefont
  {Dzurak}}]{zhao2019single}%
  \BibitemOpen
  \bibfield  {author} {\bibinfo {author} {\bibfnamefont {R.}~\bibnamefont
  {Zhao}}, \bibinfo {author} {\bibfnamefont {T.}~\bibnamefont {Tanttu}},
  \bibinfo {author} {\bibfnamefont {K.~Y.}\ \bibnamefont {Tan}}, \bibinfo
  {author} {\bibfnamefont {B.}~\bibnamefont {Hensen}}, \bibinfo {author}
  {\bibfnamefont {K.~W.}\ \bibnamefont {Chan}}, \bibinfo {author}
  {\bibfnamefont {J.}~\bibnamefont {Hwang}}, \bibinfo {author} {\bibfnamefont
  {R.}~\bibnamefont {Leon}}, \bibinfo {author} {\bibfnamefont {C.~H.}\
  \bibnamefont {Yang}}, \bibinfo {author} {\bibfnamefont {W.}~\bibnamefont
  {Gilbert}}, \bibinfo {author} {\bibfnamefont {F.}~\bibnamefont {Hudson}},
  \bibinfo {author} {\bibfnamefont {K.}~\bibnamefont {Itoh}}, \bibinfo {author}
  {\bibfnamefont {A.}~\bibnamefont {Kiselev}}, \bibinfo {author} {\bibfnamefont
  {T.}~\bibnamefont {Ladd}}, \bibinfo {author} {\bibfnamefont {A.}~\bibnamefont
  {Morello}}, \bibinfo {author} {\bibfnamefont {A.}~\bibnamefont {Laucht}}, \
  and\ \bibinfo {author} {\bibfnamefont {A.}~\bibnamefont {Dzurak}},\ }\href
  {https://doi.org/10.1038/s41467-019-13416-7} {\bibfield  {journal} {\bibinfo
  {journal} {Nature Communications}\ }\textbf {\bibinfo {volume} {10}},\
  \bibinfo {pages} {1} (\bibinfo {year} {2019})}\BibitemShut {NoStop}%
\bibitem [{\citenamefont {van~der Wiel}\ \emph {et~al.}(2002)\citenamefont
  {van~der Wiel}, \citenamefont {De~Franceschi}, \citenamefont {Elzerman},
  \citenamefont {Fujisawa}, \citenamefont {Tarucha},\ and\ \citenamefont
  {Kouwenhoven}}]{vanderWiel:2002gra}%
  \BibitemOpen
  \bibfield  {author} {\bibinfo {author} {\bibfnamefont {W.~G.}\ \bibnamefont
  {van~der Wiel}}, \bibinfo {author} {\bibfnamefont {S.}~\bibnamefont
  {De~Franceschi}}, \bibinfo {author} {\bibfnamefont {J.~M.}\ \bibnamefont
  {Elzerman}}, \bibinfo {author} {\bibfnamefont {T.}~\bibnamefont {Fujisawa}},
  \bibinfo {author} {\bibfnamefont {S.}~\bibnamefont {Tarucha}}, \ and\
  \bibinfo {author} {\bibfnamefont {L.~P.}\ \bibnamefont {Kouwenhoven}},\
  }\href {\doibase 10.1103/RevModPhys.75.1} {\bibfield  {journal} {\bibinfo
  {journal} {Rev. Mod. Phys.}\ }\textbf {\bibinfo {volume} {75}},\ \bibinfo
  {pages} {1} (\bibinfo {year} {2002})}\BibitemShut {NoStop}%
\bibitem [{\citenamefont {{Jones}}\ \emph {et~al.}(2018)\citenamefont
  {{Jones}}, \citenamefont {{Pritchett}}, \citenamefont {{Chen}}, \citenamefont
  {{Keating}}, \citenamefont {{Andrews}}, \citenamefont {{Blumoff}},
  \citenamefont {{De Lorenzo}}, \citenamefont {{Eng}}, \citenamefont {{Ha}},
  \citenamefont {{Kiselev}}, \citenamefont {{Meenehan}}, \citenamefont
  {{Merkel}}, \citenamefont {{Wright}}, \citenamefont {{Edge}}, \citenamefont
  {{Ross}}, \citenamefont {{Rakher}}, \citenamefont {{Borselli}},\ and\
  \citenamefont {{Hunter}}}]{Jones:2018vc}%
  \BibitemOpen
  \bibfield  {author} {\bibinfo {author} {\bibfnamefont {A.~M.}\ \bibnamefont
  {{Jones}}}, \bibinfo {author} {\bibfnamefont {E.~J.}\ \bibnamefont
  {{Pritchett}}}, \bibinfo {author} {\bibfnamefont {E.~H.}\ \bibnamefont
  {{Chen}}}, \bibinfo {author} {\bibfnamefont {T.~E.}\ \bibnamefont
  {{Keating}}}, \bibinfo {author} {\bibfnamefont {R.~W.}\ \bibnamefont
  {{Andrews}}}, \bibinfo {author} {\bibfnamefont {J.~Z.}\ \bibnamefont
  {{Blumoff}}}, \bibinfo {author} {\bibfnamefont {L.~A.}\ \bibnamefont {{De
  Lorenzo}}}, \bibinfo {author} {\bibfnamefont {K.}~\bibnamefont {{Eng}}},
  \bibinfo {author} {\bibfnamefont {S.~D.}\ \bibnamefont {{Ha}}}, \bibinfo
  {author} {\bibfnamefont {A.~A.}\ \bibnamefont {{Kiselev}}}, \bibinfo {author}
  {\bibfnamefont {S.~M.}\ \bibnamefont {{Meenehan}}}, \bibinfo {author}
  {\bibfnamefont {S.~T.}\ \bibnamefont {{Merkel}}}, \bibinfo {author}
  {\bibfnamefont {J.~A.}\ \bibnamefont {{Wright}}}, \bibinfo {author}
  {\bibfnamefont {L.~F.}\ \bibnamefont {{Edge}}}, \bibinfo {author}
  {\bibfnamefont {R.~S.}\ \bibnamefont {{Ross}}}, \bibinfo {author}
  {\bibfnamefont {M.~T.}\ \bibnamefont {{Rakher}}}, \bibinfo {author}
  {\bibfnamefont {M.~G.}\ \bibnamefont {{Borselli}}}, \ and\ \bibinfo {author}
  {\bibfnamefont {A.}~\bibnamefont {{Hunter}}},\ }\href@noop {} {\bibfield
  {journal} {\bibinfo  {journal} {ArXiv e-prints}\ } (\bibinfo {year}
  {2018})},\ \Eprint {http://arxiv.org/abs/1809.08320} {arXiv:1809.08320
  [quant-ph]} \BibitemShut {NoStop}%
\bibitem [{\citenamefont {Hanson}\ \emph {et~al.}(2007)\citenamefont {Hanson},
  \citenamefont {Kouwenhoven}, \citenamefont {Petta}, \citenamefont {Tarucha},\
  and\ \citenamefont {Vandersypen}}]{hanson2007spins}%
  \BibitemOpen
  \bibfield  {author} {\bibinfo {author} {\bibfnamefont {R.}~\bibnamefont
  {Hanson}}, \bibinfo {author} {\bibfnamefont {L.}~\bibnamefont {Kouwenhoven}},
  \bibinfo {author} {\bibfnamefont {J.}~\bibnamefont {Petta}}, \bibinfo
  {author} {\bibfnamefont {S.}~\bibnamefont {Tarucha}}, \ and\ \bibinfo
  {author} {\bibfnamefont {L.}~\bibnamefont {Vandersypen}},\ }\href {\doibase
  https://doi.org/10.1103/RevModPhys.79.1217} {\bibfield  {journal} {\bibinfo
  {journal} {Reviews of Modern Physics}\ }\textbf {\bibinfo {volume} {79}},\
  \bibinfo {pages} {1217} (\bibinfo {year} {2007})}\BibitemShut {NoStop}%
\bibitem [{\citenamefont {Kodera}\ \emph {et~al.}(2009)\citenamefont {Kodera},
  \citenamefont {Ono}, \citenamefont {Amaha}, \citenamefont {Arakawa},\ and\
  \citenamefont {Tarucha}}]{kodera2009pauli}%
  \BibitemOpen
  \bibfield  {author} {\bibinfo {author} {\bibfnamefont {T.}~\bibnamefont
  {Kodera}}, \bibinfo {author} {\bibfnamefont {K.}~\bibnamefont {Ono}},
  \bibinfo {author} {\bibfnamefont {S.}~\bibnamefont {Amaha}}, \bibinfo
  {author} {\bibfnamefont {Y.}~\bibnamefont {Arakawa}}, \ and\ \bibinfo
  {author} {\bibfnamefont {S.}~\bibnamefont {Tarucha}},\ }in\ \href {\doibase
  https://doi.org/10.1088/1742-6596/150/2/022043} {\emph {\bibinfo {booktitle}
  {Journal of Physics: Conference Series}}},\ Vol.\ \bibinfo {volume} {150}\
  (\bibinfo {organization} {IOP Publishing},\ \bibinfo {year} {2009})\ p.\
  \bibinfo {pages} {022043}\BibitemShut {NoStop}%
\bibitem [{\citenamefont {Chen}\ \emph {et~al.}(2017)\citenamefont {Chen},
  \citenamefont {Wang}, \citenamefont {Cao}, \citenamefont {Li}, \citenamefont
  {Xiao}, \citenamefont {Guo}, \citenamefont {Jiang}, \citenamefont {Hu},\ and\
  \citenamefont {Guo}}]{chen2017spin}%
  \BibitemOpen
  \bibfield  {author} {\bibinfo {author} {\bibfnamefont {B.-B.}\ \bibnamefont
  {Chen}}, \bibinfo {author} {\bibfnamefont {B.-C.}\ \bibnamefont {Wang}},
  \bibinfo {author} {\bibfnamefont {G.}~\bibnamefont {Cao}}, \bibinfo {author}
  {\bibfnamefont {H.-O.}\ \bibnamefont {Li}}, \bibinfo {author} {\bibfnamefont
  {M.}~\bibnamefont {Xiao}}, \bibinfo {author} {\bibfnamefont {G.-C.}\
  \bibnamefont {Guo}}, \bibinfo {author} {\bibfnamefont {H.-W.}\ \bibnamefont
  {Jiang}}, \bibinfo {author} {\bibfnamefont {X.}~\bibnamefont {Hu}}, \ and\
  \bibinfo {author} {\bibfnamefont {G.-P.}\ \bibnamefont {Guo}},\ }\href@noop
  {} {\bibfield  {journal} {\bibinfo  {journal} {Physical Review B}\ }\textbf
  {\bibinfo {volume} {95}},\ \bibinfo {pages} {035408} (\bibinfo {year}
  {2017})}\BibitemShut {NoStop}%
\bibitem [{\citenamefont {Petit}\ \emph {et~al.}(2019)\citenamefont {Petit},
  \citenamefont {Eenink}, \citenamefont {Russ}, \citenamefont {Lawrie},
  \citenamefont {Hendrickx}, \citenamefont {Clarke}, \citenamefont
  {Vandersypen},\ and\ \citenamefont {Veldhorst}}]{petit2019universal}%
  \BibitemOpen
  \bibfield  {author} {\bibinfo {author} {\bibfnamefont {L.}~\bibnamefont
  {Petit}}, \bibinfo {author} {\bibfnamefont {H.}~\bibnamefont {Eenink}},
  \bibinfo {author} {\bibfnamefont {M.}~\bibnamefont {Russ}}, \bibinfo {author}
  {\bibfnamefont {W.}~\bibnamefont {Lawrie}}, \bibinfo {author} {\bibfnamefont
  {N.}~\bibnamefont {Hendrickx}}, \bibinfo {author} {\bibfnamefont
  {J.}~\bibnamefont {Clarke}}, \bibinfo {author} {\bibfnamefont
  {L.}~\bibnamefont {Vandersypen}}, \ and\ \bibinfo {author} {\bibfnamefont
  {M.}~\bibnamefont {Veldhorst}},\ }\href {https://arxiv.org/abs/1910.05289}
  {\bibfield  {journal} {\bibinfo  {journal} {arXiv preprint arXiv:1910.05289}\
  } (\bibinfo {year} {2019})}\BibitemShut {NoStop}%
\bibitem [{\citenamefont {Yang}\ \emph {et~al.}(2019)\citenamefont {Yang},
  \citenamefont {Leon}, \citenamefont {Hwang}, \citenamefont {Saraiva},
  \citenamefont {Tanttu}, \citenamefont {Huang}, \citenamefont {Lemyre},
  \citenamefont {Chan}, \citenamefont {Tan}, \citenamefont {Hudson},
  \citenamefont {Itoh}, \citenamefont {Morello}, \citenamefont
  {Pioro-Ladri\`{e}re}, \citenamefont {Laucht},\ and\ \citenamefont
  {Dzurak}}]{yang2019silicon}%
  \BibitemOpen
  \bibfield  {author} {\bibinfo {author} {\bibfnamefont {C.}~\bibnamefont
  {Yang}}, \bibinfo {author} {\bibfnamefont {R.}~\bibnamefont {Leon}}, \bibinfo
  {author} {\bibfnamefont {J.}~\bibnamefont {Hwang}}, \bibinfo {author}
  {\bibfnamefont {A.}~\bibnamefont {Saraiva}}, \bibinfo {author} {\bibfnamefont
  {T.}~\bibnamefont {Tanttu}}, \bibinfo {author} {\bibfnamefont
  {W.}~\bibnamefont {Huang}}, \bibinfo {author} {\bibfnamefont {J.~C.}\
  \bibnamefont {Lemyre}}, \bibinfo {author} {\bibfnamefont {K.}~\bibnamefont
  {Chan}}, \bibinfo {author} {\bibfnamefont {K.}~\bibnamefont {Tan}}, \bibinfo
  {author} {\bibfnamefont {F.}~\bibnamefont {Hudson}}, \bibinfo {author}
  {\bibfnamefont {K.}~\bibnamefont {Itoh}}, \bibinfo {author} {\bibfnamefont
  {A.}~\bibnamefont {Morello}}, \bibinfo {author} {\bibfnamefont
  {M.}~\bibnamefont {Pioro-Ladri\`{e}re}}, \bibinfo {author} {\bibfnamefont
  {A.}~\bibnamefont {Laucht}}, \ and\ \bibinfo {author} {\bibfnamefont
  {A.}~\bibnamefont {Dzurak}},\ }\href@noop {} {\bibfield  {journal} {\bibinfo
  {journal} {ArXiv e-prints}\ } (\bibinfo {year} {2019})},\ \Eprint
  {http://arxiv.org/abs/1902.09126} {arXiv:1902.09126} \BibitemShut {NoStop}%
\bibitem [{\citenamefont {Leon}\ \emph {et~al.}(2020)\citenamefont {Leon},
  \citenamefont {Yang}, \citenamefont {Hwang}, \citenamefont {Lemyre},
  \citenamefont {Tanttu}, \citenamefont {Huang}, \citenamefont {Chan},
  \citenamefont {Tan}, \citenamefont {Hudson}, \citenamefont {Itoh},
  \citenamefont {Morello}, \citenamefont {Lauch}, \citenamefont
  {Pioro-Ladri\`{e}re}, \citenamefont {Saraiva},\ and\ \citenamefont
  {AS}}]{leon2020coherent}%
  \BibitemOpen
  \bibfield  {author} {\bibinfo {author} {\bibfnamefont {R.}~\bibnamefont
  {Leon}}, \bibinfo {author} {\bibfnamefont {C.~H.}\ \bibnamefont {Yang}},
  \bibinfo {author} {\bibfnamefont {J.}~\bibnamefont {Hwang}}, \bibinfo
  {author} {\bibfnamefont {J.~C.}\ \bibnamefont {Lemyre}}, \bibinfo {author}
  {\bibfnamefont {T.}~\bibnamefont {Tanttu}}, \bibinfo {author} {\bibfnamefont
  {W.}~\bibnamefont {Huang}}, \bibinfo {author} {\bibfnamefont {K.~W.}\
  \bibnamefont {Chan}}, \bibinfo {author} {\bibfnamefont {K.}~\bibnamefont
  {Tan}}, \bibinfo {author} {\bibfnamefont {F.}~\bibnamefont {Hudson}},
  \bibinfo {author} {\bibfnamefont {K.}~\bibnamefont {Itoh}}, \bibinfo {author}
  {\bibfnamefont {A.}~\bibnamefont {Morello}}, \bibinfo {author} {\bibfnamefont
  {A.}~\bibnamefont {Lauch}}, \bibinfo {author} {\bibfnamefont
  {M.}~\bibnamefont {Pioro-Ladri\`{e}re}}, \bibinfo {author} {\bibfnamefont
  {A.}~\bibnamefont {Saraiva}}, \ and\ \bibinfo {author} {\bibfnamefont
  {D.}~\bibnamefont {AS}},\ }\href@noop {} {\bibfield  {journal} {\bibinfo
  {journal} {Nature Communications}\ }\textbf {\bibinfo {volume} {11}},\
  \bibinfo {pages} {1} (\bibinfo {year} {2020})}\BibitemShut {NoStop}%
\bibitem [{\citenamefont {Huebl}\ \emph {et~al.}(2010)\citenamefont {Huebl},
  \citenamefont {Nugroho}, \citenamefont {Morello}, \citenamefont {Escott},
  \citenamefont {Eriksson}, \citenamefont {Yang}, \citenamefont {Jamieson},
  \citenamefont {Clark},\ and\ \citenamefont {Dzurak}}]{huebl2010electron}%
  \BibitemOpen
  \bibfield  {author} {\bibinfo {author} {\bibfnamefont {H.}~\bibnamefont
  {Huebl}}, \bibinfo {author} {\bibfnamefont {C.~D.}\ \bibnamefont {Nugroho}},
  \bibinfo {author} {\bibfnamefont {A.}~\bibnamefont {Morello}}, \bibinfo
  {author} {\bibfnamefont {C.~C.}\ \bibnamefont {Escott}}, \bibinfo {author}
  {\bibfnamefont {M.~A.}\ \bibnamefont {Eriksson}}, \bibinfo {author}
  {\bibfnamefont {C.}~\bibnamefont {Yang}}, \bibinfo {author} {\bibfnamefont
  {D.~N.}\ \bibnamefont {Jamieson}}, \bibinfo {author} {\bibfnamefont {R.~G.}\
  \bibnamefont {Clark}}, \ and\ \bibinfo {author} {\bibfnamefont {A.~S.}\
  \bibnamefont {Dzurak}},\ }\href {\doibase
  https://doi.org/10.1103/PhysRevB.81.235318} {\bibfield  {journal} {\bibinfo
  {journal} {Physical Review B}\ }\textbf {\bibinfo {volume} {81}},\ \bibinfo
  {pages} {235318} (\bibinfo {year} {2010})}\BibitemShut {NoStop}%
\bibitem [{\citenamefont {Sakaki}\ \emph {et~al.}(1970)\citenamefont {Sakaki},
  \citenamefont {Hoh},\ and\ \citenamefont {Sugano}}]{Sakaki:1970ii}%
  \BibitemOpen
  \bibfield  {author} {\bibinfo {author} {\bibfnamefont {H.}~\bibnamefont
  {Sakaki}}, \bibinfo {author} {\bibfnamefont {K.}~\bibnamefont {Hoh}}, \ and\
  \bibinfo {author} {\bibfnamefont {T.}~\bibnamefont {Sugano}},\ }\href
  {\doibase 10.1109/T-ED.1970.17092} {\bibfield  {journal} {\bibinfo  {journal}
  {IEEE Trans. Electron Devices}\ }\textbf {\bibinfo {volume} {17}},\ \bibinfo
  {pages} {892} (\bibinfo {year} {1970})}\BibitemShut {NoStop}%
\bibitem [{\citenamefont {Kim}\ \emph {et~al.}(2017)\citenamefont {Kim},
  \citenamefont {Tyryshkin},\ and\ \citenamefont {Lyon}}]{Kim:2017hh}%
  \BibitemOpen
  \bibfield  {author} {\bibinfo {author} {\bibfnamefont {J.~S.}\ \bibnamefont
  {Kim}}, \bibinfo {author} {\bibfnamefont {A.~M.}\ \bibnamefont {Tyryshkin}},
  \ and\ \bibinfo {author} {\bibfnamefont {S.~A.}\ \bibnamefont {Lyon}},\
  }\href {\doibase 10.1063/1.4979035} {\bibfield  {journal} {\bibinfo
  {journal} {Appl. Phys. Lett.}\ }\textbf {\bibinfo {volume} {110}},\ \bibinfo
  {pages} {123505} (\bibinfo {year} {2017})}\BibitemShut {NoStop}%
\end{thebibliography}%
\end{document}